\def\Year{%
    \the\year
}

\def\Title{%
    Investigation of the efficiency of the moving least squares method in the reconstruction of a three-dimensional surface on a supercomputer%
}

\def\Author{%
    Khabibulin Marat Ildarovich%
}

\def\SciAdvisor{%
    Nikolsky Ilya Mikhailovich%
}

\def\Position{%
    assistant professor%
}
\def\AcademicDegree{%
    Ph.D.%
}

\def\Place{%
    Moscow%
}
\def\Univer{%
    Lomonosov Moscow State University%
}
\def\Faculty{%
    Faculty of Computational Mathematics and Cybernetics%
}
\def\Department{%
    Department of Supercomputers and Quantum Information Science%
}     

\def\Status{%
    final%
}

\def\EnableSign{%
     true%
}

\documentclass[
    a4paper,
    article,
    oneside,
    onecolumn,
    12pt,
    extrafontsizes,
    openright,
    \Status
]{memoir}

\usepackage{ifthen}

\usepackage{caption}
\usepackage{subcaption}

\usepackage[T1, T2A]{fontenc}
\usepackage[utf8]{inputenc}
\usepackage[english]{babel}
\usepackage{textcomp}

\usepackage{amsthm}
\usepackage{amstext}
\usepackage{amsmath}
\usepackage{amssymb}
\usepackage{amsfonts}
\usepackage{xfrac}
\usepackage{mathtools}
\usepackage[bigdelims,vvarbb]{newtxmath}
\usepackage{upgreek}

\numberwithin{equation}{section}

\usepackage{indentfirst}
\usepackage{misccorr}

\usepackage{hyperref}
\usepackage[dvipsnames, table, hyperref]{xcolor}
\usepackage{csquotes}
\usepackage{hyphenat}

\counterwithout{section}{chapter}
\setsecnumdepth{subsection}
\maxtocdepth{subsection}

\settrimmedsize{297mm}{210mm}{*}
\settypeblocksize{620pt}{448.13pt}{*} 
\setulmarginsandblock{2cm}{2cm}{*} 
\setlrmarginsandblock{3cm}{1.5cm}{*} 
\setmarginnotes{17pt}{51pt}{\onelineskip} 
\setheadfoot{2\onelineskip}{1.5\onelineskip} 
\setheaderspaces{*}{\onelineskip}{*} 
\checkandfixthelayout

\OnehalfSpacing

\makepagestyle{standart}
\makeevenhead{standart}{}{}{}  
\makeoddhead{standart}{}{}{}

\makeevenfoot{standart}{}{\thepage}{}
\makeoddfoot{standart}{}{\thepage}{}

\aliaspagestyle{plain}{standart}

\usepackage{changepage}
\usepackage{pdflscape}
\usepackage{pdfpages}

\usepackage{tikz}
\usepackage{graphicx}
\newsubfloat{figure}

\graphicspath{{images/}}

\DeclareGraphicsExtensions{.pdf,.png,.jpg}

\usepackage{booktabs}
\usepackage{longtable}
\usepackage{tabu}

\usepackage{listings}

\usepackage[%
    ruled,
    vlined,
    algosection,
    linesnumbered,
    titlenumbered
]{algorithm2e}

\usepackage{algorithmic}

\newlength{\twless}
\newlength{\lmarg}
\usepackage[%
  backend=biber, 
  bibstyle=gost-numeric, 
  citestyle=numeric-comp, 
  language=auto, 
  sorting=none, 
  doi=true, 
  eprint=false, 
  isbn=false, 
  dashed=true, 
  url=true
]{biblatex}

\usepackage{multirow}
\newcounter{index}

\renewcommand{\epsilon}{\ensuremath{\upvarepsilon}}   
\renewcommand{\phi}{\ensuremath{\upvarphi}}
\renewcommand{\kappa}{\ensuremath{\varkappa}}
\renewcommand{\alpha}{\ensuremath{\upalpha}}
\renewcommand{\beta}{\ensuremath{\upbeta}}
\renewcommand{\gamma}{\ensuremath{\upgamma}}
\renewcommand{\delta}{\ensuremath{\updelta}}
\renewcommand{\varepsilon}{\ensuremath{\upvarepsilon}}
\renewcommand{\zeta}{\ensuremath{\upzeta}}
\renewcommand{\eta}{\ensuremath{\upeta}}
\renewcommand{\theta}{\ensuremath{\uptheta}}
\renewcommand{\vartheta}{\ensuremath{\upvartheta}}
\renewcommand{\iota}{\ensuremath{\upiota}}
\renewcommand{\kappa}{\ensuremath{\upkappa}}
\renewcommand{\lambda}{\ensuremath{\uplambda}}
\renewcommand{\mu}{\ensuremath{\upmu}}
\renewcommand{\nu}{\ensuremath{\upnu}}
\renewcommand{\xi}{\ensuremath{\upxi}}
\renewcommand{\pi}{\ensuremath{\uppi}}
\renewcommand{\varpi}{\ensuremath{\upvarpi}}
\renewcommand{\rho}{\ensuremath{\uprho}}
\renewcommand{\varrho}{\ensuremath{\upvarrho}}
\renewcommand{\sigma}{\ensuremath{\upsigma}}
\renewcommand{\varsigma}{\ensuremath{\upvarsigma}}
\renewcommand{\tau}{\ensuremath{\uptau}}
\renewcommand{\upsilon}{\ensuremath{\upupsilon}}
\renewcommand{\varphi}{\ensuremath{\upvarphi}}
\renewcommand{\chi}{\ensuremath{\upchi}}
\renewcommand{\psi}{\ensuremath{\uppsi}}
\renewcommand{\omega}{\ensuremath{\upomega}}

\begin{document}

\makepagestyle{titlepagestyle}
\makeevenhead{titlepagestyle}{}{}{}  
\makeoddhead{titlepagestyle}{}{}{}

\makeevenfoot{titlepagestyle}{}{}{}
\makeoddfoot{titlepagestyle}{}{\Place, \Year}{}
\thispagestyle{titlepagestyle}
\begin{SingleSpace}
\begin{center}
\includegraphics{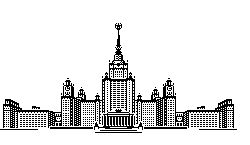}\\
{\small
\MakeUppercase{\Univer}\\
\MakeUppercase{\Faculty}\\
\MakeUppercase{\Department}\\[4\baselineskip]
}
\Author\\
\textbf{
  \Title\\[2\baselineskip]
}
\end{center}

\vspace*{2\baselineskip}

\noindent
\ifx\SciAdvisor\empty
\else
    \begin{minipage}{0.4\textwidth}
    \phantom{MMMM}
    \end{minipage}
    \begin{minipage}{0.58\textwidth}
            \SingleSpacing
            \textbf{Scientific director:}\\
            \Position
            \ifx\AcademicDegree\empty
                \\%
            \else
                , \AcademicDegree\\%
            \SciAdvisor
            \fi
    \end{minipage}

    \def\True{true}
    \ifx\EnableSign\True
    \vspace*{4\baselineskip}
    \begin{minipage}{0.4\textwidth}
    \phantom{MMMM}
    \end{minipage}
    \begin{minipage}{0.58\textwidth}
    \end{minipage}
    \fi
\fi

\end{SingleSpace}

    \clearpage
    \tableofcontents*
    \clearpage

    
\phantomsection
\section*{Annotation}
\addcontentsline{toc}{section}{Annotation}
The work is devoted to the problem of surface reconstruction - the classical problem of processing a digital representation of a scanned physical form. There are a large number of methods that reconstruct a surface from a set of discrete points. This paper considers the moving least squares method, which is distinguished by the fact that it calculates a point representation of the resulting surface and is resistant to noisy input data. A parallel version of the algorithm is proposed, designed for a distributed computing system. Computational experiments carried out on a high-performance Polus cluster showed good scalability of the proposed implementation of the moving least squares method.
\clearpage

\section*{Introduction} 
\addcontentsline{toc}{section}{Introduction}
Currently, the field of geometric modeling and the construction of 3D models using point clouds obtained from laser sensors is actively developing. One of the basic tasks of geometric modeling is surface reconstruction from a point cloud. Surface reconstruction methods have a wide range of applications:
\begin{itemize}
    \item surface reconstruction is used in face recognition;
    \item surface reconstruction is used to build three-dimensional models of objects using sensors such as lidar and depth camera;
    \item in robotics, in localization problems and the construction of three-dimensional maps based on point clouds from sensors, reconstruction is an important stage of the task;
    \item surface reconstruction is used in medicine;
    \item surface reconstruction is used in machine vision;
\end{itemize}

\begin{figure}[h]
    \centering
    \includegraphics[width=0.32\textwidth, height=5cm]{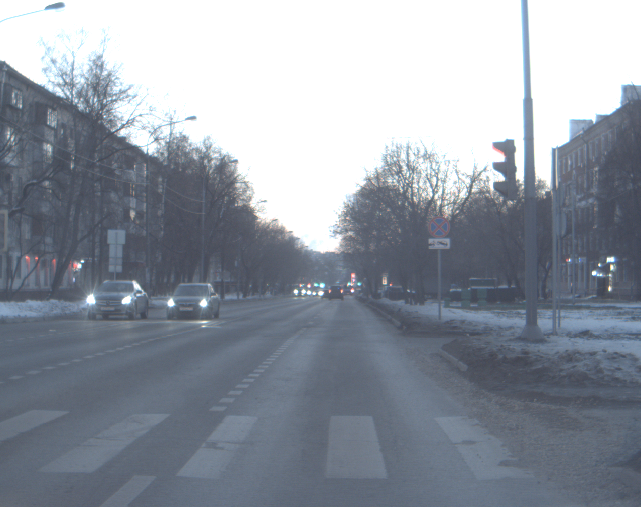}
    \includegraphics[width=0.32\textwidth, height=5cm]{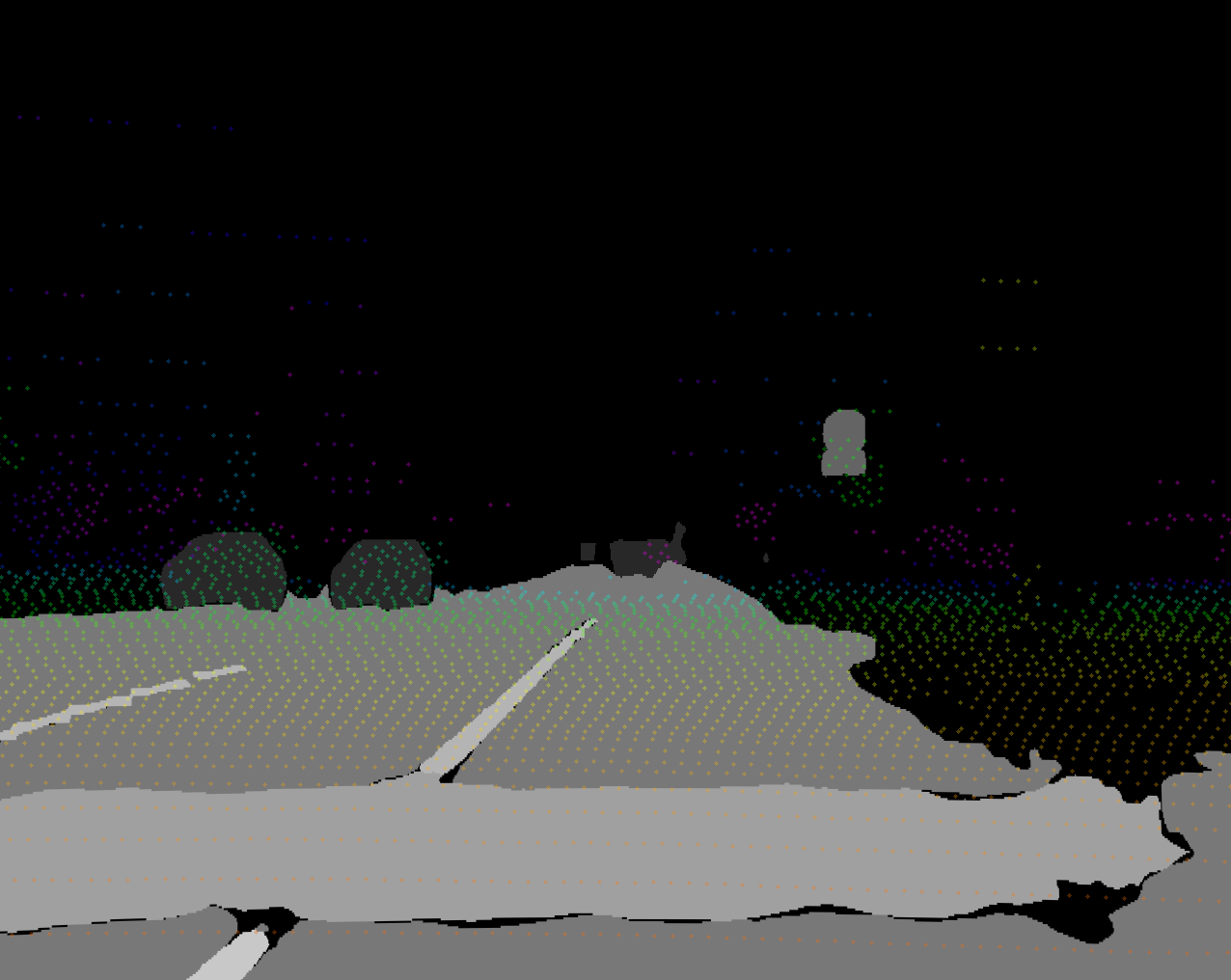}
    \includegraphics[width=0.32\textwidth, height=5cm]{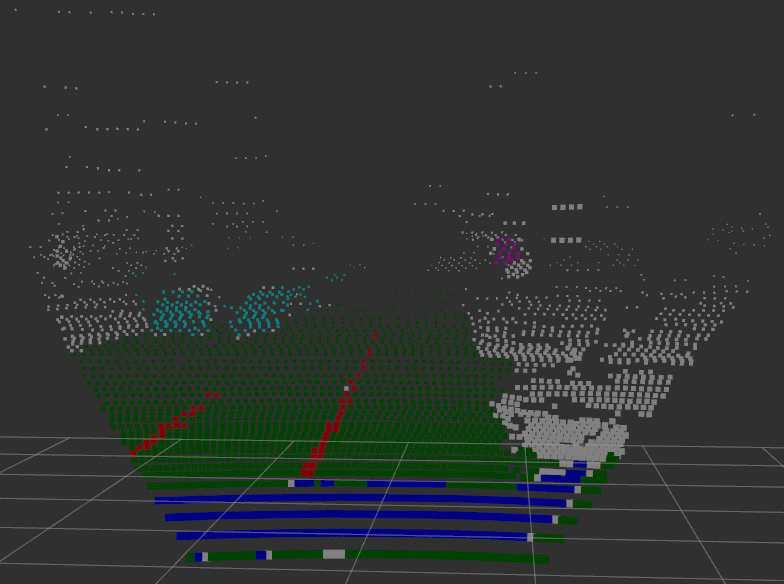}
    \caption{Point cloud segmentation using a segmented image mask and one-to-one correspondence between lidar and camera data.}
    \label{fig:-2}
\end{figure}

The use of surface reconstruction methods in 3D cartography is very interesting and promising. To construct an accurate map requires long and expensive labor of markers. One of the directions in automating this process is the following solution. A bug file is recorded with data from the lidar and camera (the scan from the lidar and the image from the camera are synchronized in time) when the car moves along the street. A segmenting neural network is applied to the camera images (see Fig. \ref{fig:-2}), a reconstruction algorithm is applied to the lidar scans (sampling is increased, gaps are filled, noise and outliers are removed), then the point cloud is projected onto the mask, obtained by solving a neural network (this is possible thanks to internal calibration of the camera (projection matrix) and external calibration (transition from the lidar coordinate system to the camera coordinate system (shift and rotation vector))). Using a one-to-one correspondence between data from the lidar and the camera, the point cloud is also segmented (see Fig. \ref{fig:-2}).

\newpage

\begin{figure}[h]
    \centering
    \includegraphics[width=0.8\textwidth, height=5cm]{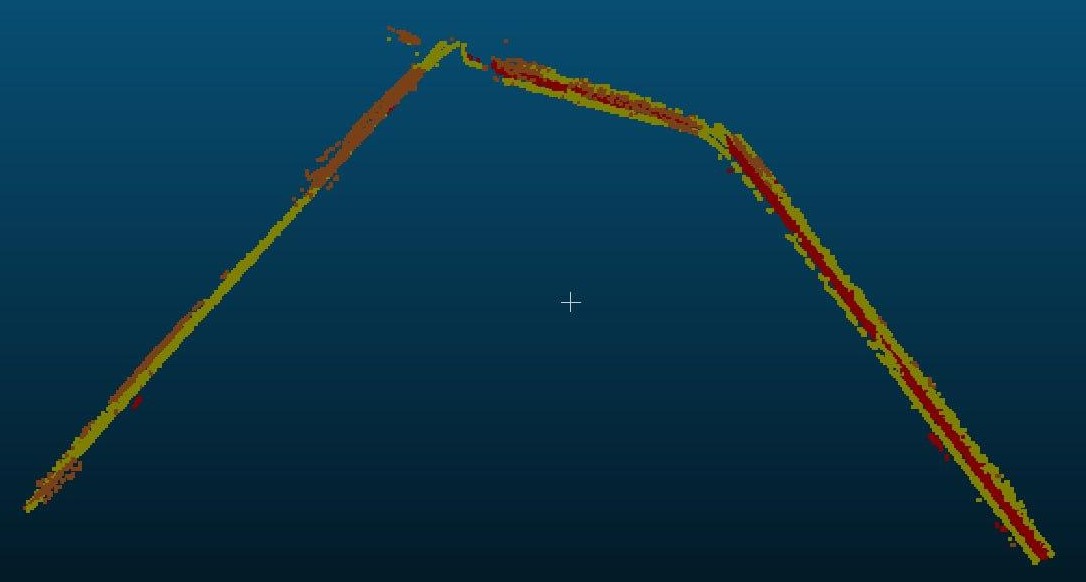}
    \caption{Stitched segmented scans from lidar.}
    \label{fig:-1}
\end{figure}

Scans from the lidar are merged into one continuous scan (see Fig. \ref{fig:-1})

\begin{figure}[h]
    \centering
    \includegraphics[width=0.8\textwidth, height=5cm]{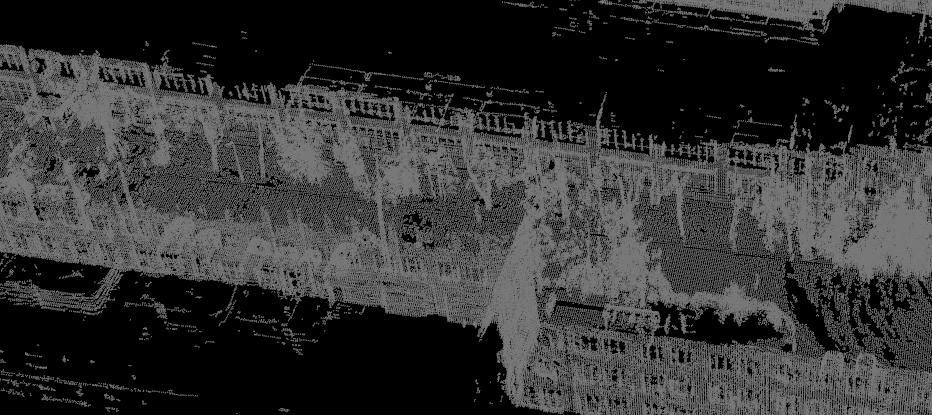}
    \caption{Stitched scans of street houses.}
    \label{fig:street}
\end{figure}

Reconstruction algorithms are also applied to buildings from the street to define them with a polygonal mesh (see Fig. \ref{fig:street}).

Due to the sharp increase in the availability of lidar sensors, today they can be seen on some models of popular phones (iphone 12 pro, etc.). This has contributed to the emergence of numerous startups related to the production of orthopedic shoes based on an individual cast of the foot obtained through scanning and subsequent reconstruction of the surface of the foot. When making professional sports equipment (such as skates, helmets), three-dimensional scanning of the relevant parts of the body is used to obtain an impression.

In the first section of the work, a review of the theory is carried out, surface reconstruction methods are considered that represent the surface with a mesh, as well as a set of highly discretized points. Algorithms for Delaunay triangulation, surfaces of algebraic points, and the method of moving least squares are considered.

The second section describes the features of the parallel implementation of the MLS algorithm. The algorithm has a large parallelism resource due to its locality of calculations, as well as the absence of the problem of combining surface pieces due to the point representation of the surface (as opposed to a grid representation). The section presents a diagram of the parallel algorithm, as well as graphs of information dependence of parallelized loops. When calculating a polynomial defining MLS surface points for a point in space (query point), you need to search through the entire point cloud to find elements included in a sphere of radius \textbf{r} (method parameter) and centered at the query point. This is a standard spatial indexing task that can be accelerated by using various data structures. Most often, a k-d tree ~\cite{Russell} is used for this task, since it is easy to build and run queries on it. The algorithm takes advantage of the fact that the input data, such as a lidar scan, when split, represents continuous parts of the surface. Thus, when constructing k-d trees, they contain points from continuous parts of surfaces, which eliminates the need to use k-d tree data from neighboring processes. In the case when the order of points in the cloud is chaotic, you can use one of the methods for sorting points, for example, reordering them in Morton order ~\cite{MORTON}. The section also presents the order of computational complexity of the various stages of the algorithm. The nonlinear computational complexity of the k-d tree construction stage leads to the fact that an increase in the number of trees associated with an increase in the number of processes leads to a nonlinear decrease in the number of operations performed on the process, which has a beneficial effect on the efficiency of the algorithm.

The third section contains the results of the parallel algorithm on a supercomputer, their analysis, conclusions and recommendations.

The problem of determining a surface from a set of points has been actively studied for many years. Despite the proliferation of surface reconstruction methods, many aspects of the problem remain open. Most surface reconstruction algorithms represent the surface as a mesh. But there are also algorithms that represent a surface as a highly discretized set of points.

Some of the main difficulties in the reconstruction process are the complexity of the shape and noise. The focus of the work is on a surface restoration algorithm based on the least squares method, which is why the algorithm is called the Moving Least Squares method. The algorithm is called moving because it iteratively moves through a set of points. This algorithm has various variations. The most used option is the MLS projection operator. The MLS projection operator ~\cite{LEVIN} has proven to be a powerful method for surface reconstruction.
The MLS method allows us to achieve a simple and efficient representation of a surface by a set of points. MLS surfaces are now widely used in the processing and rendering of point-sampled models and are increasingly used as a standard point-set definition of surfaces. The computation of points on a surface is local, resulting in a non-standard technique that can handle any set of points. Besides the surface reconstruction by the smoothing function, the moving least squares method has many other advantages, such as the inherent ability to handle noisy input data, the ease of computing differential geometric surface properties (e.g., normals, curvature).

In differential geometry, a smooth surface is characterized by the existence of smooth local mappings at any point. MLS provides such a mapping on a local domain of definition, a local coordinate system. And although formally, it is possible to define the MLS surface with implicit functions, in practice these functions are approximated by a set of points from these functions, with discretization in the order of image resolution. This is due to the large number of polynomials, the difficulty of recalculating them as implicit functions to a general coordinate system and specifying the domain of their definition, and other ambiguous problems. The work ~\cite{LEVIN} shows that the error of such an approximation is limited and depends on the distance between points. Thus, it is possible to ensure a predetermined accuracy of surface approximation. There are many different variants of the MLS algorithm, but they mainly differ in the choice of local domain.

Improving surface quality or improving the visual quality of a geometric surface is subjective. The claim that one method produces the best surface quality may vary from person to person. For this reason, it is necessary to establish numerical measures to compare the impact of surface restoration algorithms on surface quality.
In geometric modeling, a quantity called average geometric deviation is used to describe the level of surface distortion. The work also calculates the standard deviation.


\textbf{The practical significance} of the work lies in the possibility of using its results for a wide range of applications that require surface reconstruction from a point cloud. This work can help the designer choose the optimal surface reconstruction method.

\section*{Goal of work and tasks}
\addcontentsline{toc}{section}{Goal of work and tasks}
\textbf{The goal} of this work is to develop a parallel method for surface restoration based on the least squares method on a distributed memory supercomputer, allowing to achieve optimal results in both scalability and quality of surface restoration.

\textbf{Tasks} of the work are:
\begin{itemize}
     \item Development of a parallel algorithm for surface restoration using the moving least squares method on distributed memory;
     \item Writing a hybrid surface restoration program on distributed memory using the MPI and openMP libraries;
     \item Testing the developed algorithm and assessing its efficiency, acceleration, scalability, as well as the quality of the reconstructed surface by standard deviation and average geometric deviation from the reference surface;
\end{itemize}
    \section{Literature review and research problem statement}
\subsection{Surface reconstruction based on Delaunay triangulation}
In modern computer graphics, the vector-polygon model is the most common. It is used in computer-aided design systems, computer game development environments, geographic information systems, CAD, etc.

Surface reconstruction methods based on Delaunay triangulation are the most widely known among polygonal methods. The main distinguishing feature is the definition of a polygonal mesh with triangles forming a graph, which has the following conditions: the edges of the graph do not intersect and the graph has the maximum number of edges, taking into account this condition, the circumcircle for any face of the graph does not contain vertices with the exception of vertices belonging to the face ~\cite{Скворцов}. 

\begin{figure}[h]
    \centering
    \includegraphics[scale=0.7]{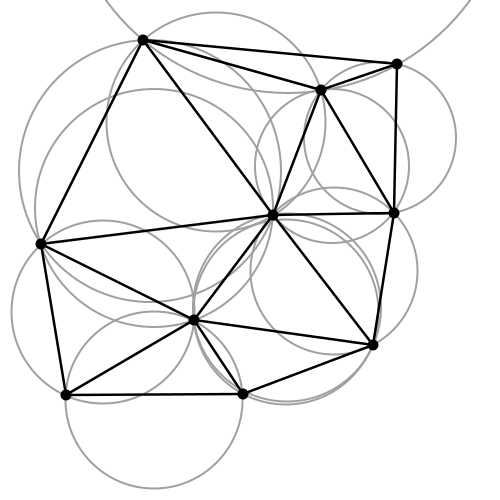}
    \caption{Execution of the Delaunay triangulation invariant}
    \label{fig:delauney}
\end{figure}

It is worth noting that the triangulation algorithms themselves do not change the position of points from the input point cloud. This is a weakness of the algorithm in terms of surface reconstruction from noisy input data. It is also worth noting the information dependence of the algorithms at each subsequent iteration on the calculations obtained in the previous operation associated with preserving the invariant of fulfilling the condition that the graph is a Delaunay triangulation. Splitting the input point cloud into segments leads to subsequent difficulties in connecting graph segments while fulfilling the same ~\cite{Tamal} invariant.
\subsection{Locally optimal projection (LOP) operator}
This method is based on the Weisfeld algorithm for solving the Fermat-Weber point location problem, also known as L1 multidimensional median. It is a statistical tool that has traditionally been used around the world for multivariate nonparametric point samples to obtain a good representative for a large number of samples in the presence of noise and outliers. This problem was first formulated by Weber in ~\cite{WEBER} under the name problem of determining the optimal location. The task was to find the optimal location for the industrial site, minimizing the cost of access. In statistics, the problem is known as L1 median ~\cite{BROWN, SMALL}.

The Fermat-Weber (global) point location problem is considered as a spatial median, since, being limited to the one-dimensional case, it coincides with the one-dimensional median and inherits some of its properties in the multidimensional formulation.

Reconstruction using the projection operator has the important benefit of identifying consistent geometry from data points and providing a constructive means to upsample it.
The parameter-free locally optimal projection operator uses a more primitive projection mechanism, but since it is not based on local 2D parameterization, it is more robust and works well in complex scenarios. Additionally, if the data points are taken locally from a smooth surface, the operator provides a second-order approximation, resulting in a plausible approximation of the selected surface.

The LOP operator has two immediate functions: first, it can be used as a preprocessing step for any other higher order reconstruction method (e.g. RBF). LOP can be applied to raw scanned data to create a clean data set, as a means to effectively reduce noise and outliers, and to facilitate the determination of local surface orientation and topology. Secondly, it can be used to refine a given data set.

For a set of data points $P = \{p_j\}_{j\in J} \subset \mathbf R^{3}$, LOP projects an arbitrary set of points $X^{(0)} = \{x_i^{( 0)} \} _{i \in I} \subset \mathbf R^{3}$ to the set $P$, where $I$, $J$ denote sets of indices. The set of projected points $Q = \{q_i\}_{i\in I}$ is defined so that it minimizes the sum of weighted distances to points P with respect to radial weights centered on the same set of points Q. Moreover, points Q must not being too close to each other. This structure induces the definition of the sought points Q as a solution to a fixed point equation
$$Q = G(Q),$$
где
$$G(C) = argmin_{X = \{x_i\}_{i \in I}} \{E_1(X,P,C) + E_2(X,C)\},$$
$$E_1(X,P,C) = \sum_{i \in I} \sum_{j \in J}\parallel x_i - p_j \parallel \theta(\parallel c_i - p_j \parallel), $$ 
$$E_2(X, C) = \sum _{i^{'} \in I} \lambda_{i^{'}}\sum_{i \in I \setminus\{i^{'}\}} \eta(\parallel x_{i^{'}}- c_i  \parallel)\theta(\parallel c_{i^{'}} - c_i \parallel)$$

Here $\theta(r)$ is a rapidly decreasing smooth weight function with a compact reference radius $h$, which determines the size of the radius of influence, $\eta(r)$ is another decreasing function that penalizes $x_{i^{'}}$ for coming too close to other points, and $\{\lambda_i\}_{i \in I}$ are balancing terms, which are denoted by $\mathbf \land$. In a nutshell, the term $E_1$ forces the projected points $Q$ to approximate the geometry of $P$, while the term $E_2$ aims to preserve a fair distribution of the points $Q$. Correct values of $\mathbf\land$ can guarantee the degree of second-order approximation of the LOP operator, provided that the data is sampled from the surface $C^{2}$.

\subsection{Radial basis functions (RBFs)}
\begin{figure}[h]
     \centering
     \includegraphics[scale=0.5]{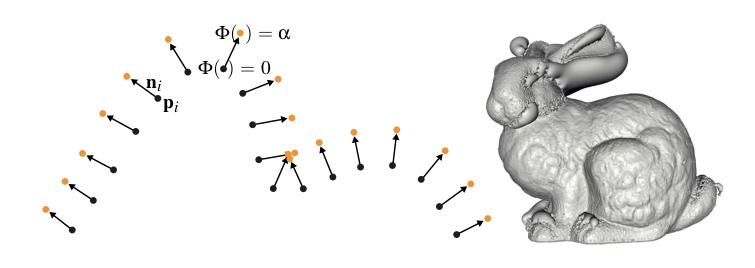}
     \caption{(left) For RBFs, the scalar field to be optimized must be estimated to be zero at the sample points $\Phi(\mathbf{p_i}) = 0$, while for off-surface constraints $\Phi(\mathbf{p_i} + \alpha\mathbf{n_i}) = \alpha;$ this choice is appropriate since signed distance functions almost everywhere have a unit gradient norm. The cluster of off-surface samples shows how carefully you need to set constraints in areas of high curvature. (right) Surface reconstructed with RBFs typically has severe geometric and topological artifacts when inconsistent external constraints are provided.}
     \label{fig:4}
\end{figure}

Radial basis functions are a well-known method for interpolating scattered data. Given a set of points with given function values, RBFs reproduce functions containing a high degree of smoothness through a linear combination of radially symmetric basis functions. For surface reconstruction, the ~\cite{CARR} method constructs the surface by finding a signed scalar field defined in terms of RBFs whose set of zero levels represents the surface. In particular, they use globally supported basis functions $\phi : R^{+} \rightarrow R$. The implicit function $\Phi$ can then be expressed as:
$$\Phi(\mathbf{x}) = g(\mathbf{x}) + \sum_j\lambda_j\phi(\parallel \mathbf{x} - \mathbf{q_j} \parallel), $$
where $g(x)$ denotes a (globally supported) low-degree polynomial and the basis functions are concentrated at the nodes $\mathbf{q_j} \in R^{3} $. The unknown coefficients ${\lambda}_j$ are found by specifying interpolation constraints on the value of the function $\theta$ at $\mathbf{p_i} \in P;$ see Fig. \ref{fig:4}. Off-surface constraints are necessary to avoid the trivial solution $f (\mathbf{x}) = 0$ for $\mathbf{x} \in R^{3}$. Positive (resp. negative) constraints are set for points displaced at point $\mathbf{p_i}$ along $\mathbf{n_i}$ in the positive (resp. negative) direction. Interpolation is performed by combining on- and off-surface constraint points as a set of node centers $\mathbf{q_{j}} $. The coefficients $\mathbf{{\lambda}_i}$ are found using a dense linear system with n unknowns, efficiently calculated using fast multipole methods ~\cite{CARR}. The advantage of using globally supported basis functions for surface reconstruction is that the resulting implicit function is globally smooth. Therefore, RBFs can be effective in creating impervious surfaces in the presence of uneven sampling and missing data. However, when the input data contains moderate noise, determining the correct placement of off-surface points can be challenging (see Fig. \ref{fig:4} right).

\subsection{Moving least squares method}
The procedure for determining a surface using the least squares method was introduced by Levin ~\cite{LEVIN}.
Let the points $p_i \in R^{3}, i \in \{1, . . . , N\}$, taken from
surface S (possibly with measurement noise). The goal is to project a point $r \in R^{3}$ near S onto a two-dimensional surface SP that approximates $p_i$. The MLS procedure is motivated by differential geometry, namely that a surface can be locally approximated by a function. The algorithm is called moving because it iteratively moves through a set of points. The point at which the iteration is located is called the query point. For the query point r (see Fig. \ref{fig:0}), the local plane H is calculated using the least squares method for points falling in the vicinity of the radius R (algorithm parameter)

Reference plane:
  Local plane $ H = \{x \mid \langle n, x \rangle - D = 0, x \in R^{3}\}, n \in R^{3}, \parallel n \parallel = 1 $ is calculated so as to minimize the local weighted sum of squared distances of points $p_i$ to the plane (see Fig. \ref{fig:0}). The weights corresponding to $p_i$ are defined as a function of the distance from $p_i$ to the projection of r onto the H plane, rather than the distance to r. Suppose q is the projection of r onto H, then H is found by local minimization
  \begin{equation}
      \sum_{i = 1}^{N}(\langle n, p_i \rangle - D)^{2} \theta(\parallel p_i - q \parallel)
      \label{eq:ref1}
  \end{equation}

where $\theta$ is a smooth monotonically decreasing function, positive over the entire space. Assuming $q = r + tn$ for some $t \in R$, the equation \ref{eq:ref1} can be rewritten as:
  $$\sum_{i = 1}^{N}(\langle n, p_i - r - tn \rangle)^{2} \theta(\parallel p_i - r - tn \parallel)$$
 
   The operator $Q(r) = q = r + tn$ is defined as the local minimum of the equation with the smallest t and the local tangent plane H near r, respectively. The local reference region is then defined by an orthonormal coordinate system on H, such that q is the origin of this system. Then l is calculated
    \section{Development of a parallel surface reconstruction algorithm}
\subsection{Proposed parallel method}
\begin{figure}[h]
   \centering
   \subbottom[$np = 1$]{%
     \includegraphics[width=0.4\textwidth, height=7cm]{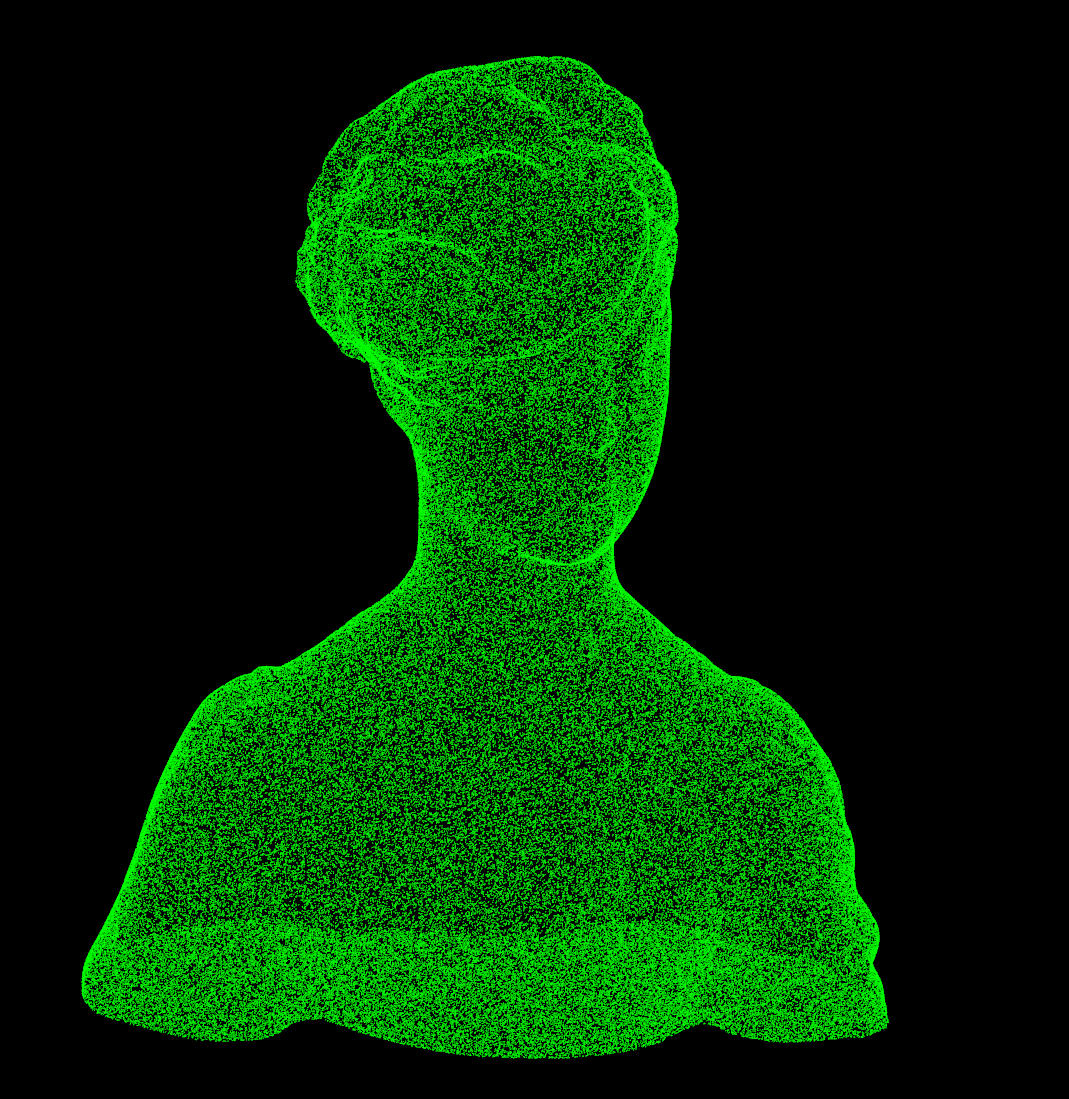}}
   \subbottom[$np = 4$]{%
     \includegraphics[width=0.58\textwidth, height=7cm]{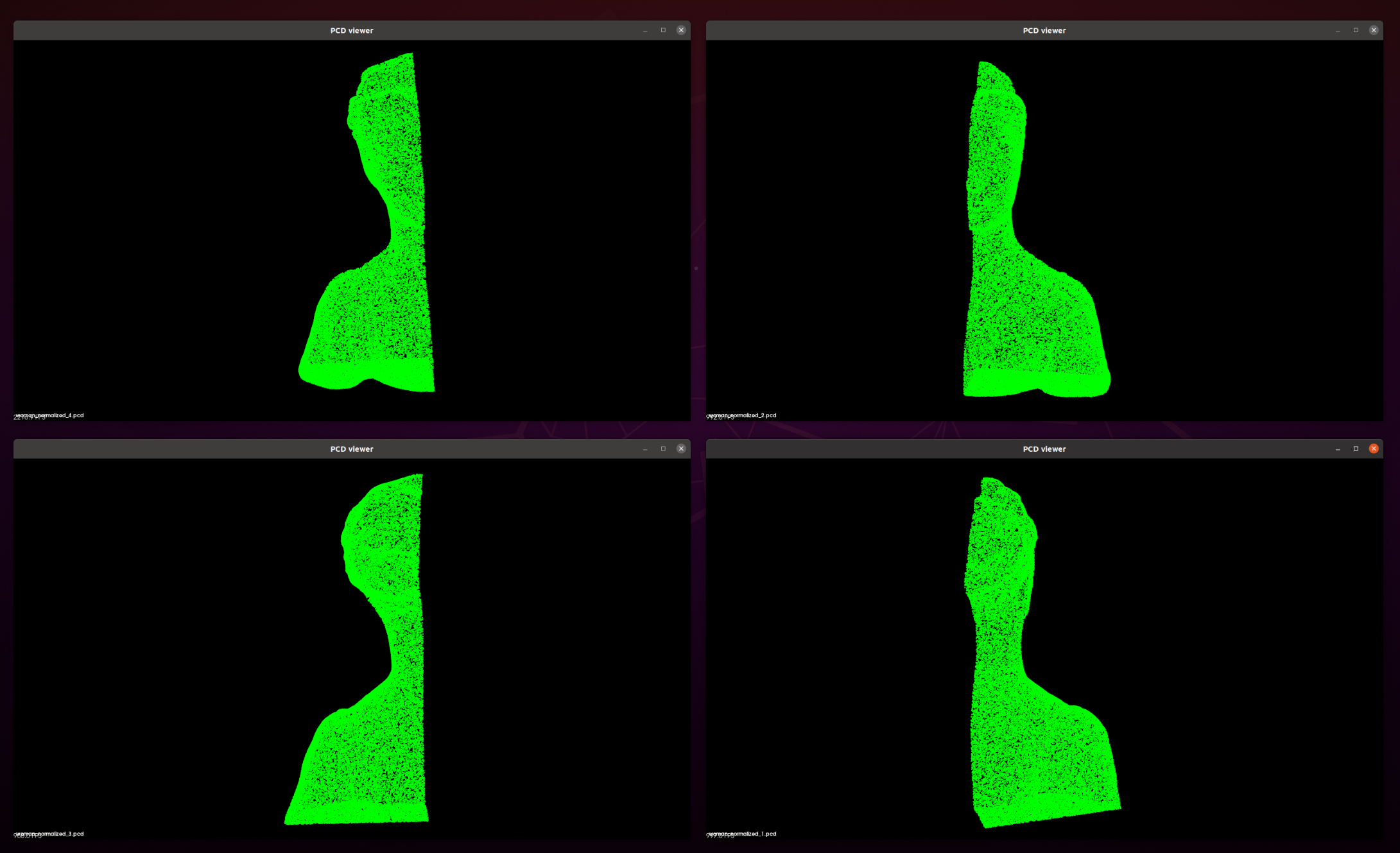}}
   \caption{Distribution of point cloud segments by process}
   \label{fig:decomposition}
\end{figure}

A parallel version of the modified MLS algorithm using MPI is described in Algorithm 1. The algorithm assumes that the point cloud is uniformly distributed across all \ref{fig:decomposition} processes. Therefore, the part of P that is locally accessible in process u is denoted by $P^{(u)}$. $P_l^{u}$, $P_r^{u}$ denote the left and right boundaries of parts of the point cloud. They are sequentially obtained from neighboring processes by exchanges in a ring topology. No additional communications are required, and the rest of the calculations are performed locally. Looping through the local point cloud $P^{(u)}$, the MLS projection procedure is performed: first, a local reference plane H is created for point $p_j$. The projection of $p_j$ onto H defines the origin of coordinates q. Then a local polynomial approximation g of the heights $f_j$ of points $p_j$ over H is calculated. The projection of $p_j$ onto g is the result of the MLS algorithm. \\*
\textbf{Algorithm 1} Parallel moving least squares method with MPI and OpenMP \\*
\textbf{Input:} set of points $P = \{p_i\}$ $i = 1..n$ \\*
\textbf{Output:} surface represented by a set of points \\*
1: \textbf{for each} process u \textbf{do} \\*
2: \quad $P^{(u)} = read(P)$ // each process reads its own segment of the point cloud $P^{(u)} = \{p_j\}$ $j = 1..m$ \\*
3: \quad $P\_l^{(u)} = send\_recv(P\_r^{(u-1)})$ // getting the left border\\*
4: \quad $P\_r^{(u)} = send\_recv(P\_l^{(u+1)})$ // getting the right border\\*
5: \quad\textbf{pragma omp parallel for} \\*
6: \quad\textbf{for each} point $j = 1..m $ \textbf{do}\\*
7: \quad\quad$H = generate\_plane(p_j)$ \\*
8: \quad\quad$g = generate\_local\_polynomial\_approximation(H)$ \\*
9: \quad\quad$result\_point = project\_on\_polynom(p_j, polynom)$ \\*
10: \quad\textbf{end for} \\*

\begin{figure}[h]
     \centering
     \includegraphics[width=0.7\textwidth, height=5cm]{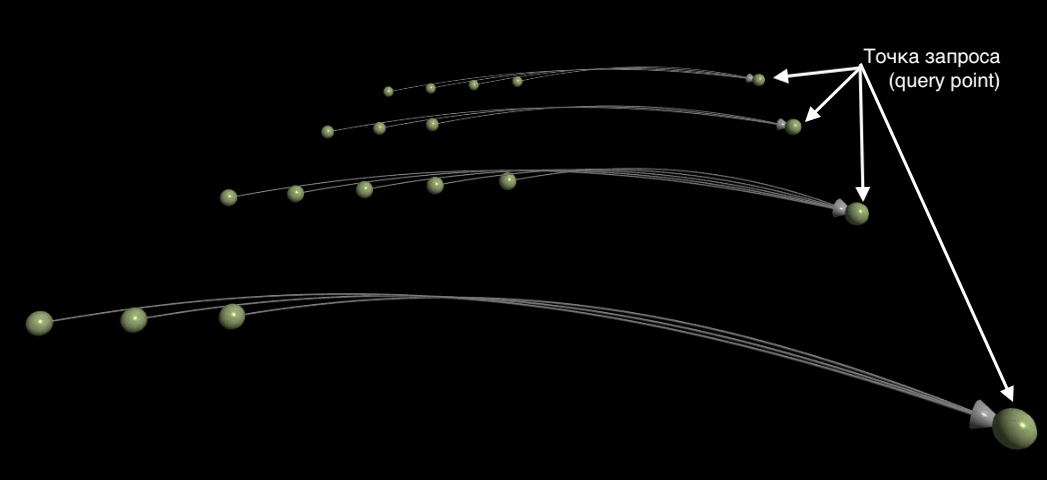}
     \caption{MLS information dependency graph.}
     \label{fig:information}
\end{figure}

When studying the information structure of the algorithm, an information dependence graph \ref{fig:information} was constructed. The graph shows the absence of information dependence between the query points, so the iterations through the loop from the local set of points $P^{(u)}$ were parallelized using OpenMP (line 5 of Algorithm 1).

\begin{figure}[h]
     \centering
     \includegraphics[scale=0.9]{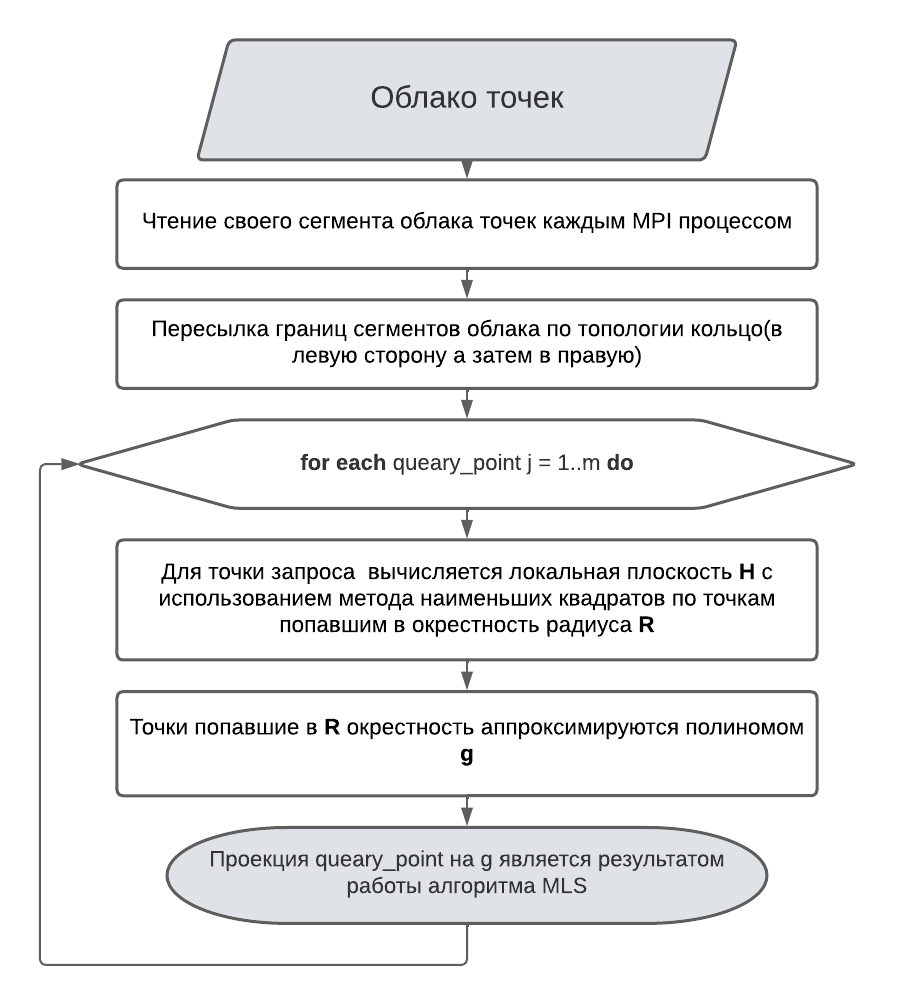}
     \caption{Block diagram of parallel implementation of the algorithm}
     \label{fig:mesh1}
\end{figure}

\clearpage

\subsection{Computational complexity}
To effectively search for points falling in the R neighborhood, the k-d tree data structure is used. Computational complexity of constructing a k-d-tree of order: $O(n*k*log(n))$. Here k = 3 and has a dimension value. An unbalanced k-d tree can be constructed in $O(n(k+log(n)))$. We have n points, each inserted at logarithmic complexity. The term $O(nk)$ has the following meaning: Before constructing the tree, the minimum and maximum are found for each dimension for subsequent uniform partitioning. In the case of a balanced tree, preliminary sorting is used over all dimensions with a complexity of the order of $O(nlog n)$.
Finding points in a neighborhood of R has a complexity of order $O(log n)$.
The least squares method for finding a plane has a complexity of order $O(C^2*m)$ $ (C = 4)$. Here C has the value of the number of parameters, and is equal to 4 since we are looking for a plane.
Interpolation by a polynomial has a complexity of the order of $O(m^2)$, where m is the number of points in the R neighborhood.

\begin{table}[h]
\centering
\begin{tabular}{||p{7cm}||p{9.3cm}||}
\hline
Algorithm stage & complexity\\
\hline\hline
Construction of a k-d-tree & $O(n \cdot k \cdot log(n))$ (Unbalanced $O(n(k+log(n)))$ ) \\
\hline
Search for points in the neighborhood of R & $O(log n)$ \\
\hline
Least squares method & $O(C^2 \cdot m) (C = 4)$\\
\hline
Interpolation by polynomial & $O(m^2)$ \\
\hline
Projection of a point onto a polynomial & $O(1)$ \\
\hline
\end{tabular}
\caption{Computational complexity of the main stages of the algorithm}
\label{table:complexity}
\end{table}

The final depreciation computational complexity of the algorithm:

$O(n \cdot k \cdot log(n))$ + n $\cdot$ ($O(log n)$ + $O(C^2 \cdot m)$ + $O(m^2)$ + $O(1)$)
    \section{Results of computational methods}

\subsection{Characteristics of a computing system for conducting experiments}

System Specification:

  Polus is a parallel computing system consisting of 5 computing nodes. (the first computing node is assigned the functions of a frontend node)
 
\noindent Main characteristics of each node:
\begin{itemize}
     \item 2 ten-core IBM POWER8 processors (each core has 8 threads) 160 threads total
     \item Total RAM 256 GB (node 5 has 1024 GB RAM) with ECC control
     \item 2 x 1 TB 2.5” 7K RPM SATA HDD
     \item 2 x NVIDIA Tesla P100 GPU, 16Gb, NVLink
     \item 1 port 100 GB/s
\end{itemize}

\noindent Cluster performance (Tflop/s): 55.84 (peak), 40.39 (Linpack) \\

\subsection{Description of experiments}
Let us now consider the question of the efficiency of the algorithm. To do this, recall that
Each parallel algorithm is evaluated based on two parameters: speedup $S_p$ and
efficiency $E_p$ , which are determined by the formulas:
$$S_p = {{t_1} \over {t_p}},$$ $$E_p = {S_p \over p} * 100\%$$
where $t_1$ is the time to solve the original problem on one processor, $t_p$ is the time
solving the original problem using a parallel algorithm on p processors.

\begin{figure}[h]
     \centering
     \includegraphics[width=0.32\textwidth, height=5cm]{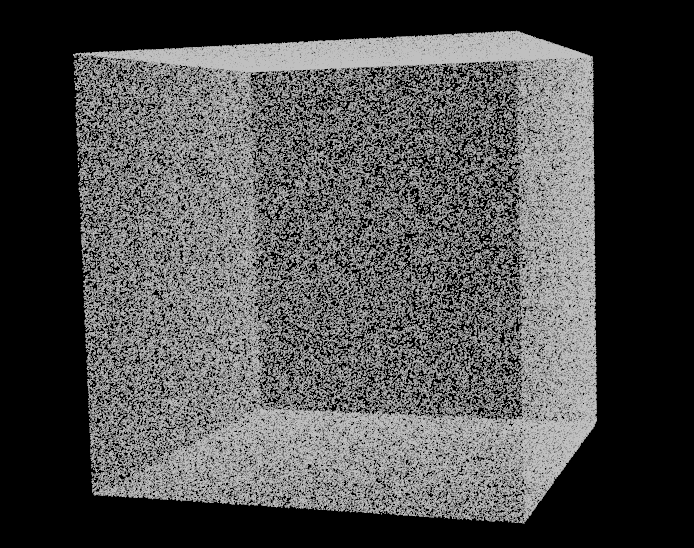}
     \includegraphics[width=0.32\textwidth, height=5cm]{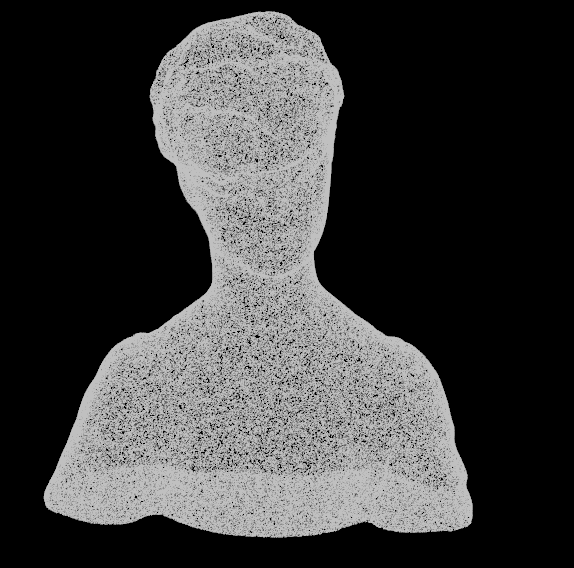}
     \includegraphics[width=0.32\textwidth, height=5cm]{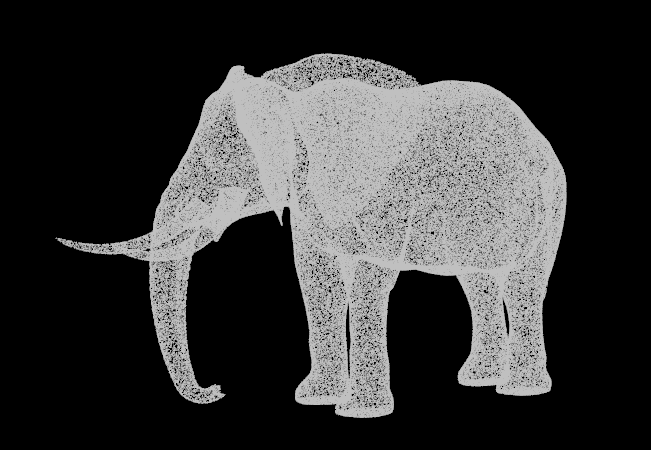}
     \includegraphics[width=0.32\textwidth, height=5cm]{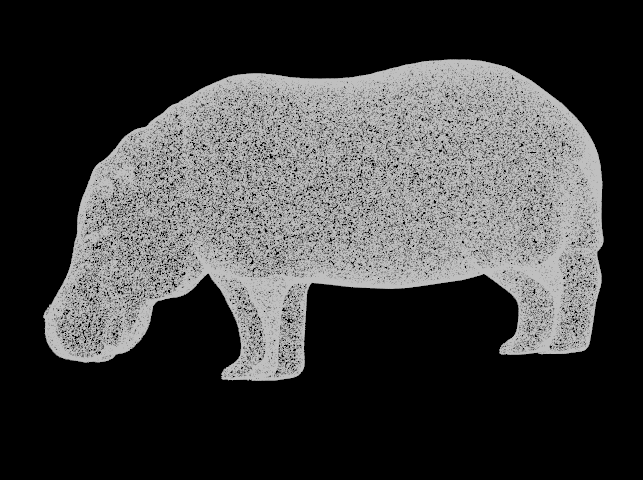}
     \includegraphics[width=0.32\textwidth, height=5cm]{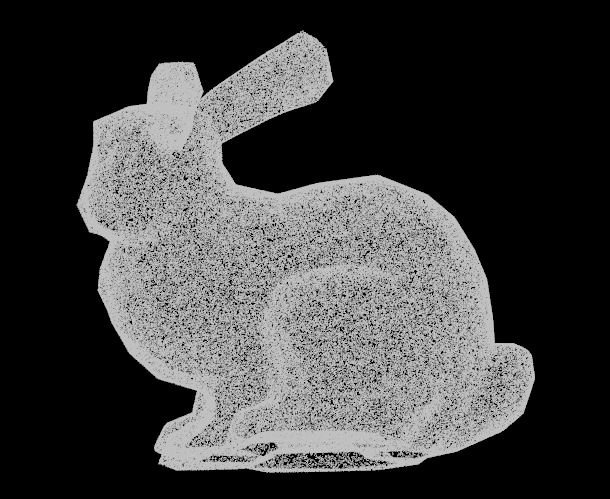}
     \includegraphics[width=0.32\textwidth, height=5cm]{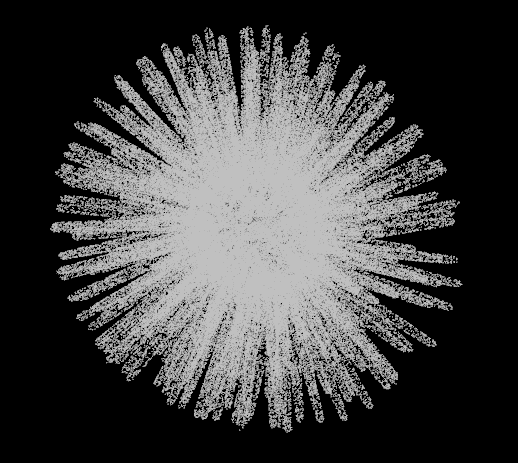}
     \caption{Point clouds taken from polygonal models. 200,000 random points were taken from each model.}
     \label{fig:point cloud models}
\end{figure}

To evaluate the performance of the algorithm in terms of the quality of surface reconstruction, the following experiments were carried out:\\
Polygonal models of varying complexity were taken, ranging from a regular cube to a sea urchin (see Fig. \ref{fig:polygonal models}). All models were pre-normalized as follows: shifted by the center of mass to the origin of coordinates, all vertices of the model were scaled so that the maximum deviation from the origin of coordinates was less than 1. These polygonal models will subsequently be called reference.\\
200,000 random points were taken from the polygonal models (see Fig. \ref{fig:point cloud models}).\\
The point cloud from the reference model was noisy with additive Gaussian noise with different standard deviations. This point cloud was subsequently used as input to the MLS algorithm.\\
The MLS algorithm with different parameters $\bold{R}$ was applied to a noisy point cloud. The surface reconstructed by the MLS algorithm, represented by a set of points, was compared with the reference model.
The average deviation of the restored surface from the reference model, as well as the standard deviation were calculated (Tables \ref{table:1}, \ref{table:2}, \ref{table:3}, \ref{table:4}, \ref {table:5},\ref{table:6}). According to research results, the MLS algorithm copes well with noisy data, but to achieve optimal results, you need to carefully select the radius parameter of the algorithm. This is especially true for models with complex shapes. A radius that is too small can lead to data loss due to the inability to find the local plane, while a radius that is too large can lead to blurring of sharp surface contours.

\begin{figure}[h]
   \centering
     \includegraphics[width=0.8\textwidth]{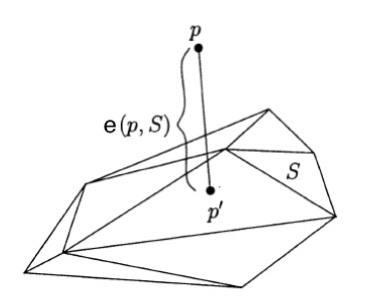}
   \caption{The deviation $e_i(p, S)$ is the distance between a point and the surface S. Point $p^{'}$ is the closest point to the surface S.}
\end{figure}

\clearpage
\subsection{Pictures, tables, graphs}

\begin{table}[h]
\begin{tabular}{|c|c|c|c|c|}
     \hline
     $R$ & Number of MPI processes & Running time (s) & Speedup & Efficiency \\
     \hline
     0.0008 & 1 & 601.91 & 1 & 100.0 \\
     0.0008 & 2 & 356.632 & 1.69 & 84.39 \\
     0.0008 & 4 & 210.948 & 2.85 & 71.33 \\
     0.0008 & 8 & 124.986 & 4.82 & 60.2 \\
     0.0008 & 16 & 73.742 & 8.16 & 51.01 \\
     0.0008 & 32 & 43.434 & 13.86 & 43.31 \\
     \hline
     0.0012 & 1 & 1070.74 & 1 & 100.0 \\
     0.0012 & 2 & 615.675 & 1.74 & 86.96 \\
     0.0012 & 4 & 342.008 & 3.13 & 78.27 \\
     0.0012 & 8 & 193.405 & 5.54 & 69.2 \\
     0.0012 & 16 & 110.338 & 9.7 & 60.65\\
     0.0012 & 32 & 63.003 & 17.0 & 53.11 \\
     \hline
     0.0016 & 1 & 1959.78 & 1 & 100.0 \\
     0.0016 & 2 & 1093.557 & 1.79 & 89.61 \\
     0.0016 & 4 & 579.585 & 3.38 & 84.53 \\
     0.0016 & 8 & 310.948 & 6.3 & 78.78 \\
     0.0016 & 16 & 168.845 & 11.61 & 72.54 \\
     0.0016 & 32 & 90.501 & 21.65 & 67.67 \\
     \hline
\end{tabular}
\caption{Results of running the program for various values of the parameter R. A cloud of 25,000,000 points was supplied as input.}
\label{table:1}
\end{table}

The graphs show measurements for various values of the parameter R when the number of MPI processes changes. The efficiency graph shows that with a larger radius, the efficiency decreases more slowly. This is due to an increase in the number of local computations with the same costs of communication between processes.

\begin{figure}[h]
     \centering
     \includegraphics[scale=0.6]{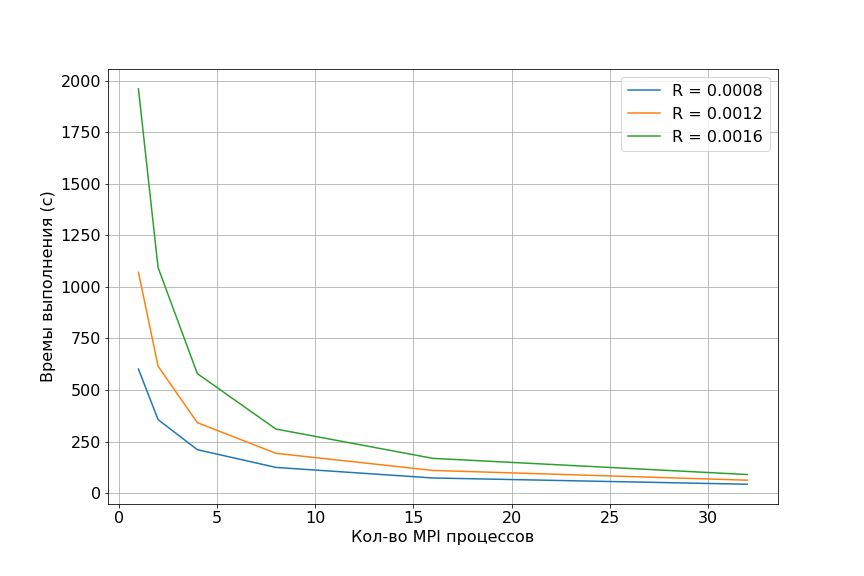}
     \caption{Working time (s) on n processes n = 1...32}
     \label{fig:mesh1}
\end{figure}

\begin{figure}[h]
     \centering
     \includegraphics[scale=0.6]{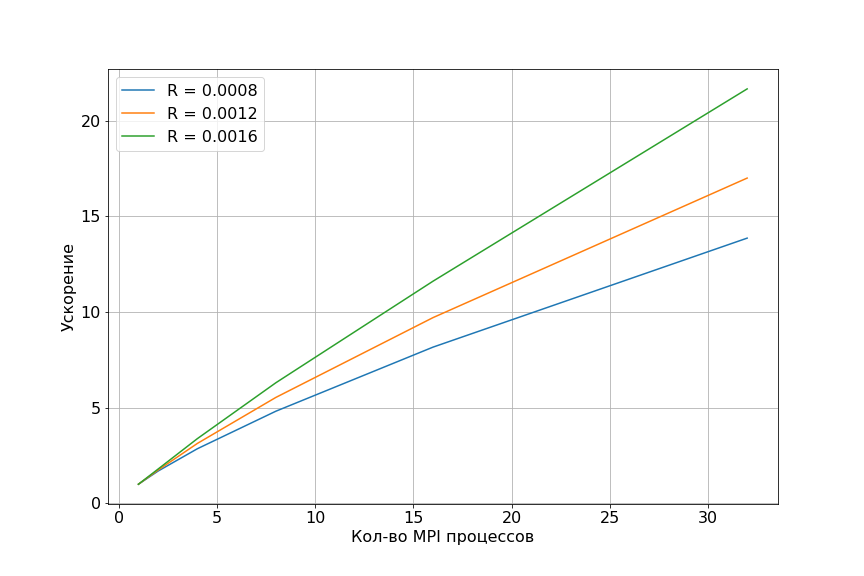}
     \caption{Acceleration}
     \label{fig:mesh1}
\end{figure}

\begin{figure}[h]
     \centering
     \includegraphics[scale=0.6]{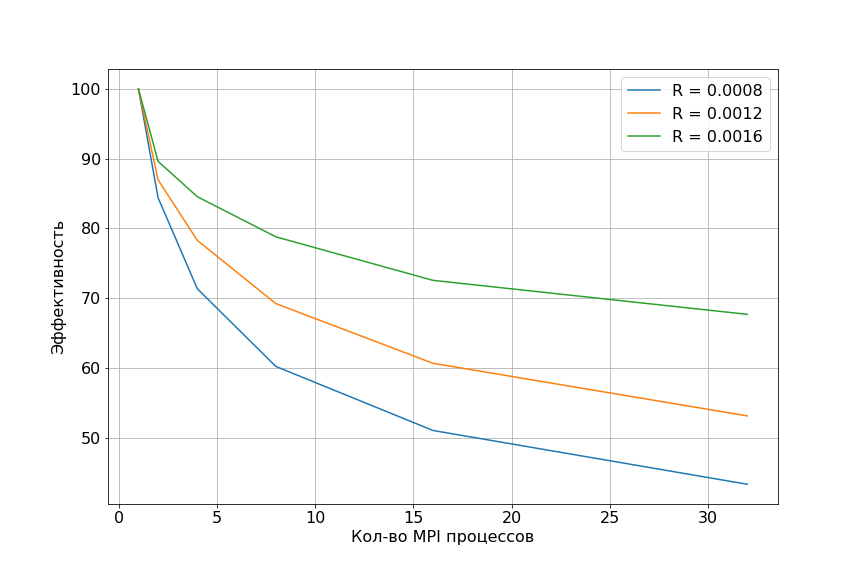}
     \caption{Efficiency}
     \label{fig:mesh1}
\end{figure}

\begin{table}[h]
\begin{tabular}{|c|c|c|c|c|c|}
     \hline
     $R$ & Number of MPI processes & Number of OpenMP threads & Running time (s) \\
     \hline\hline
     0.0008 & 1 & 1 & 601.91 \\
     0.0008 & 2 & 1 & 356.632 \\
     0.0008 & 4 & 1 & 210.948 \\
     0.0008 & 8 & 1 & 124.986 \\
     \hline
     0.0008 & 1 & 2 & 340.348 \\
     0.0008 & 2 & 2 & 201.095 \\
     0.0008 & 4 & 2 & 117.203 \\
     0.0008 & 8 & 2 & 68.377 \\
     \hline
     0.0008 & 1 & 4 & 183.182 \\
     0.0008 & 2 & 4 & 108.107 \\
     0.0008 & 4 & 4 & 63.591 \\
     0.0008 & 8 & 4 & 37.494 \\
     \hline\hline
     0.0012 & 1 & 1 & 1070.74 \\
     0.0012 & 2 & 1 & 615.675 \\
     0.0012 & 4 & 1 & 342.008 \\
     0.0012 & 8 & 1 & 193.405 \\
     \hline
     0.0012 & 1 & 2 & 602.795 \\
     0.0012 & 2 & 2 & 340.879 \\
     0.0012 & 4 & 2 & 187.004 \\
     0.0012 & 8 & 2 & 104.702 \\
     \hline
     0.0012 & 1 & 4 & 310.09 \\
     0.0012 & 2 & 4 & 176.728 \\
     0.0012 & 4 & 4 & 97.626 \\
     0.0012 & 8 & 4 & 55.051 \\
     \hline\hline
     0.0016 & 1 & 1 & 1959.78 \\
     0.0016 & 2 & 1 & 1093.557 \\
     0.0016 & 4 & 1 & 579.585 \\
     0.0016 & 8 & 1 & 310.948 \\
     \hline
     0.0016 & 1 & 2 & 1080.297 \\
     0.0016 & 2 & 2 & 598.387 \\
     0.0016 & 4 & 2 & 312.359 \\
     0.0016 & 8 & 2 & 166.638 \\
     \hline
     0.0016 & 1 & 4 & 558.458 \\
     0.0016 & 2 & 4 & 310.34 \\
     0.0016 & 4 & 4 & 163.279 \\
     0.0016 & 8 & 4 & 87.274 \\
    
     \hline
\end{tabular}
\caption{Results of running the hybrid program for various values of the parameter R. A cloud of 25,000,000 points was supplied as input.}
\label{table:1}
\end{table}

\clearpage
The following graph shows running time measurements for various values of the R parameter when changing the number of mpi processes using openMP. As a result, the hybrid program turned out to be more effective than a pure MPI program.

\begin{figure}[h]
     \centering
     \includegraphics[scale=0.6]{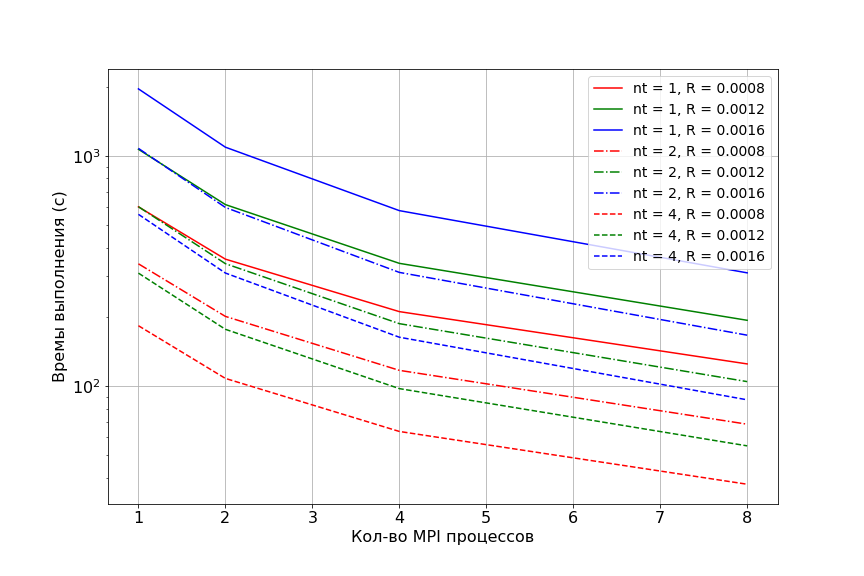}
     \caption{Running time of the hybrid program (MPI + OpenMP)}
     \label{fig:mesh1}
\end{figure}

\clearpage
All polygonal models were normalized so that the results on different models were comparable. A polygonal model consists of a list of vertices and connections between them. To normalize a polygonal model, it is enough to normalize the cloud of vertex points. The connections between the vertices remain unchanged.

\begin{figure}[h]
     \centering
     \subbottom[Cube]{\includegraphics[width=0.32\textwidth, height=5cm]{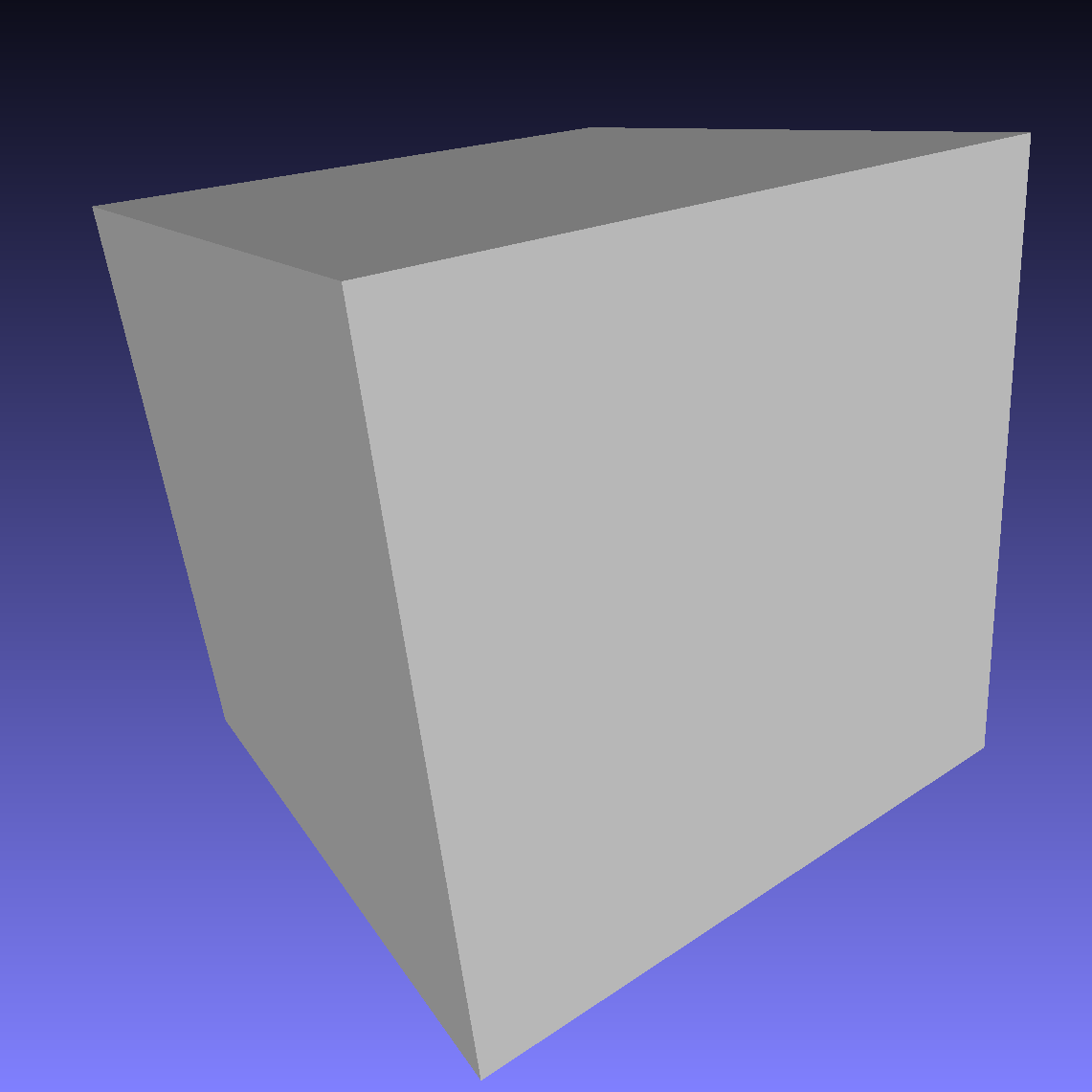}}
     \subbottom[Woman]{\includegraphics[width=0.32\textwidth, height=5cm]{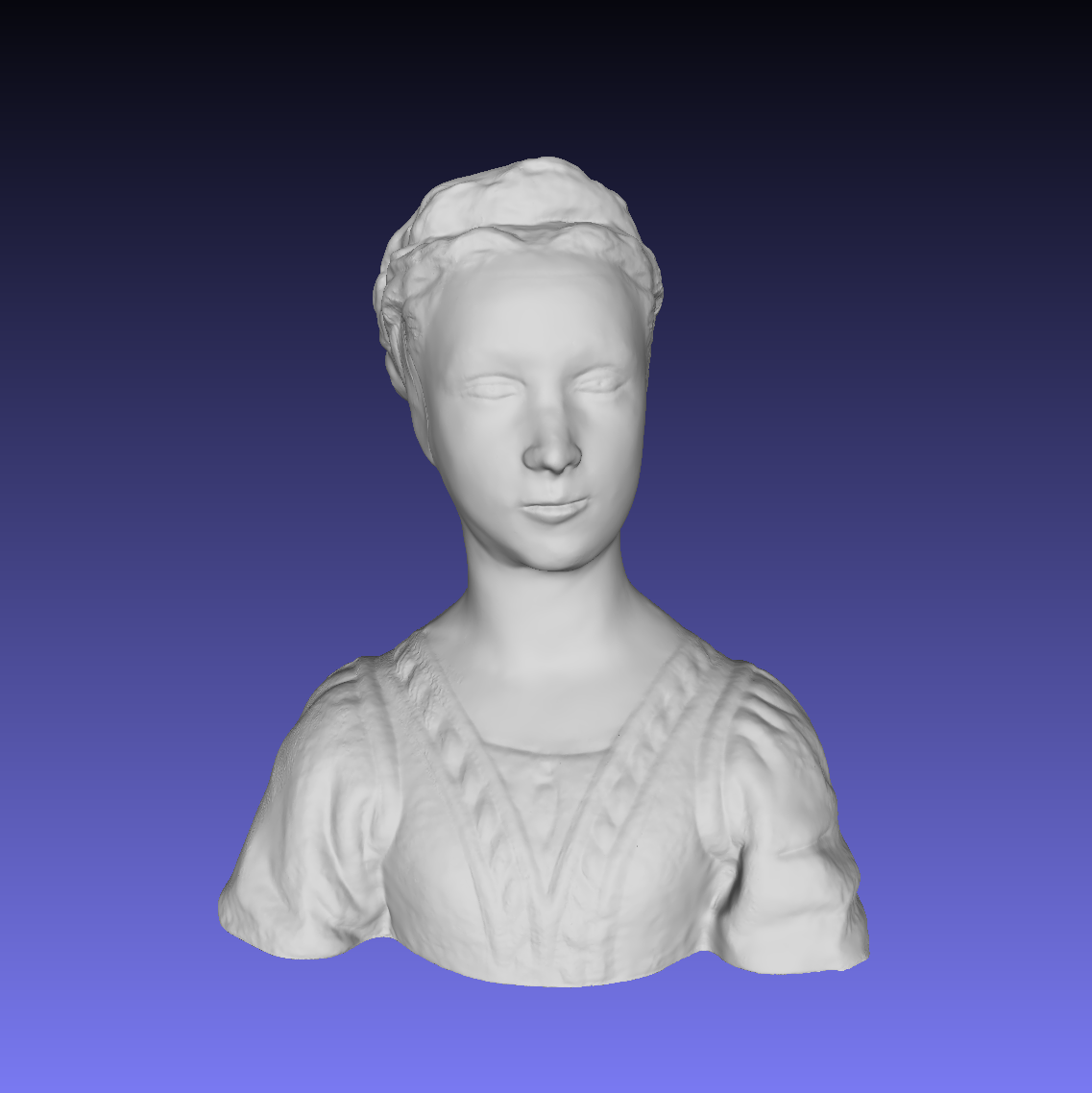}}
     \subbottom[Elephant]{\includegraphics[width=0.32\textwidth, height=5cm]{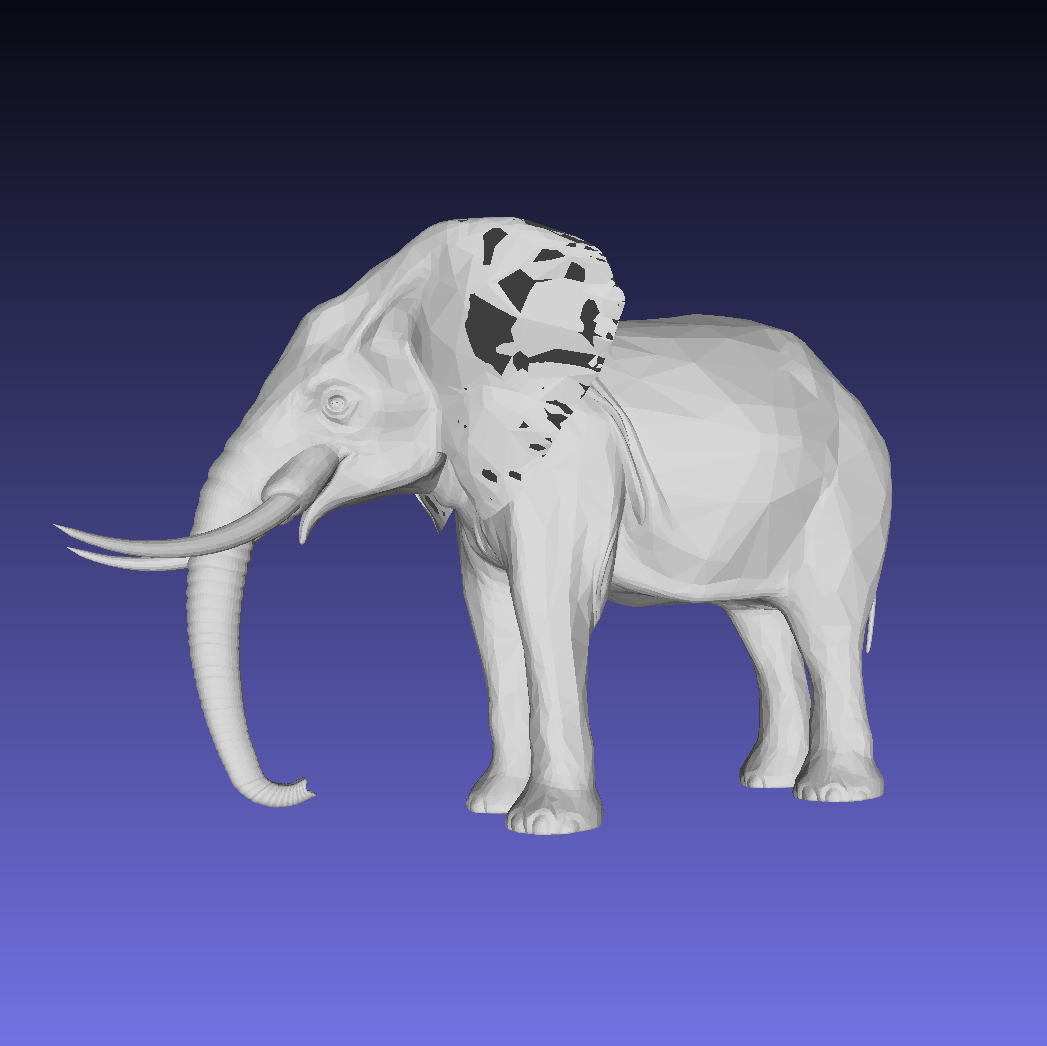}}
     \subbottom[Hippo]{\includegraphics[width=0.32\textwidth, height=5cm]{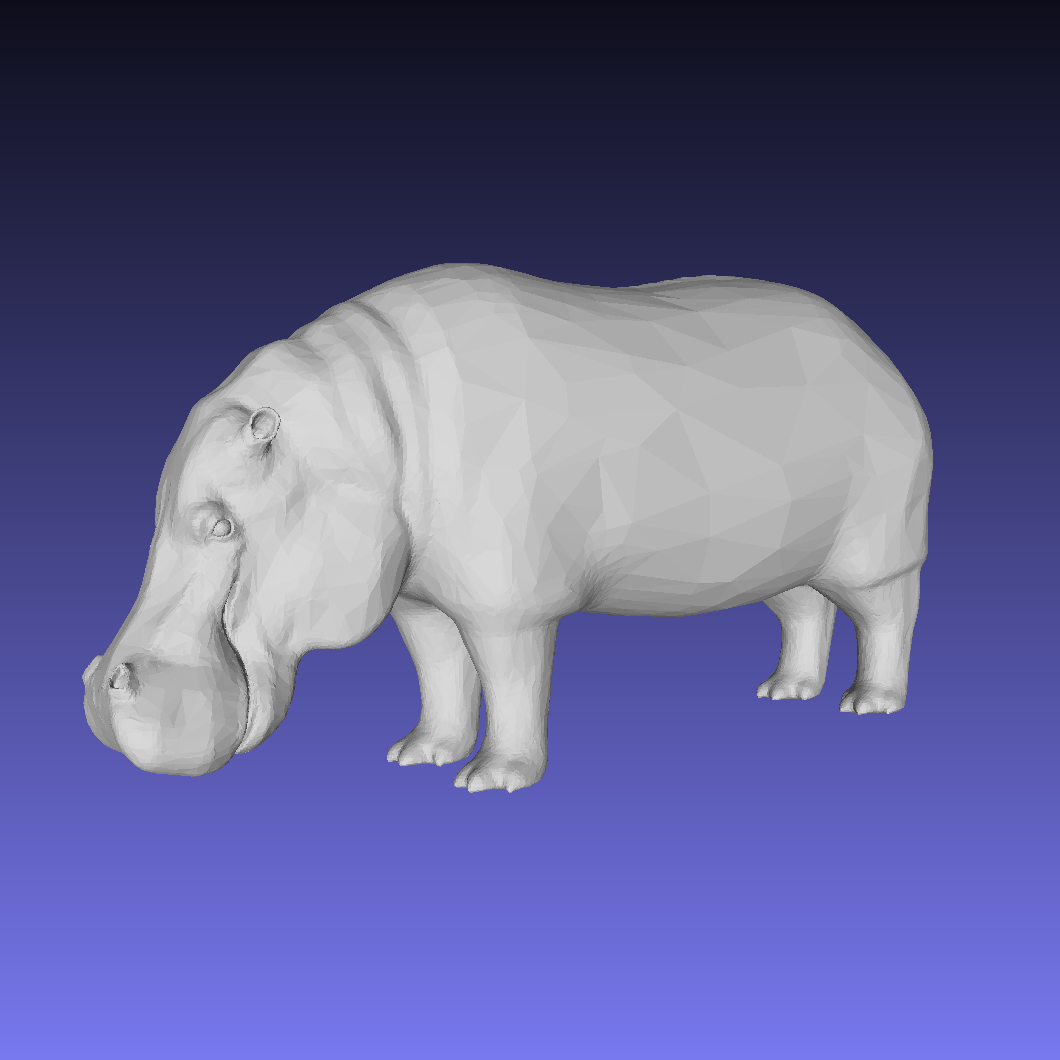}}
     \subbottom[Bunny]{\includegraphics[width=0.32\textwidth, height=5cm]{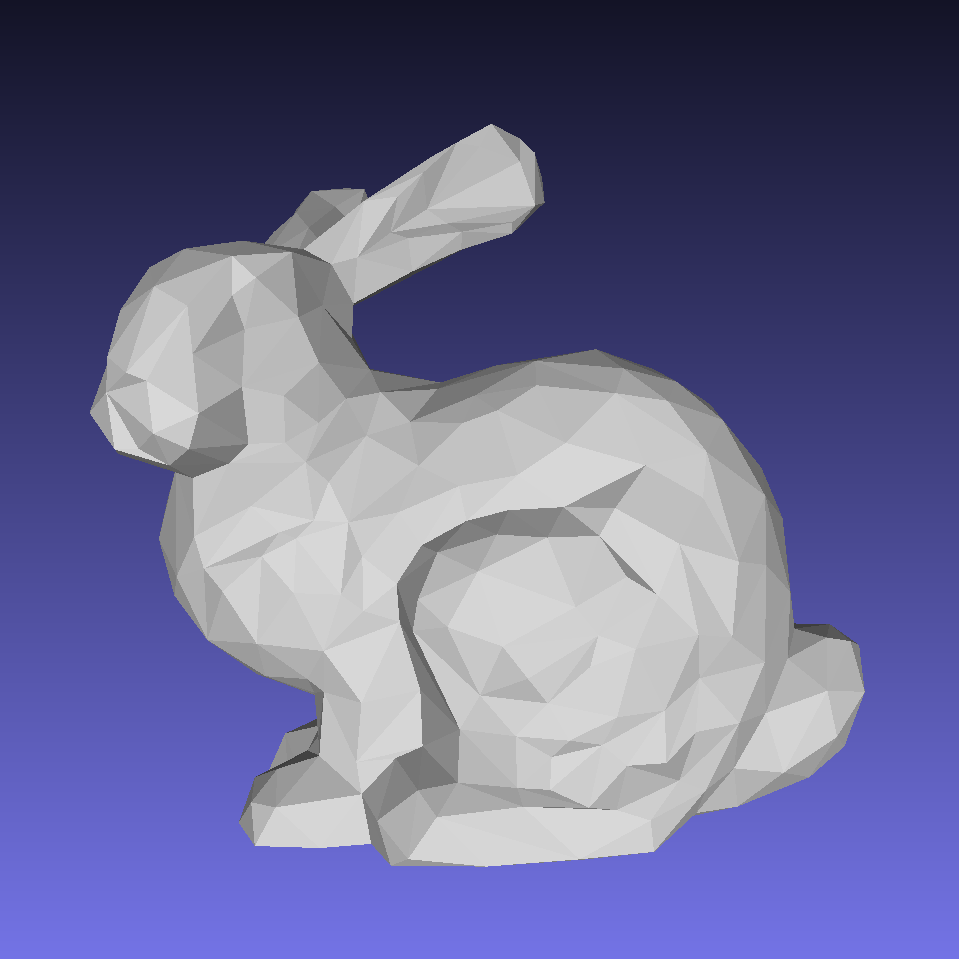}}
     \subbottom[Sea Urchin]{\includegraphics[width=0.32\textwidth, height=5cm]{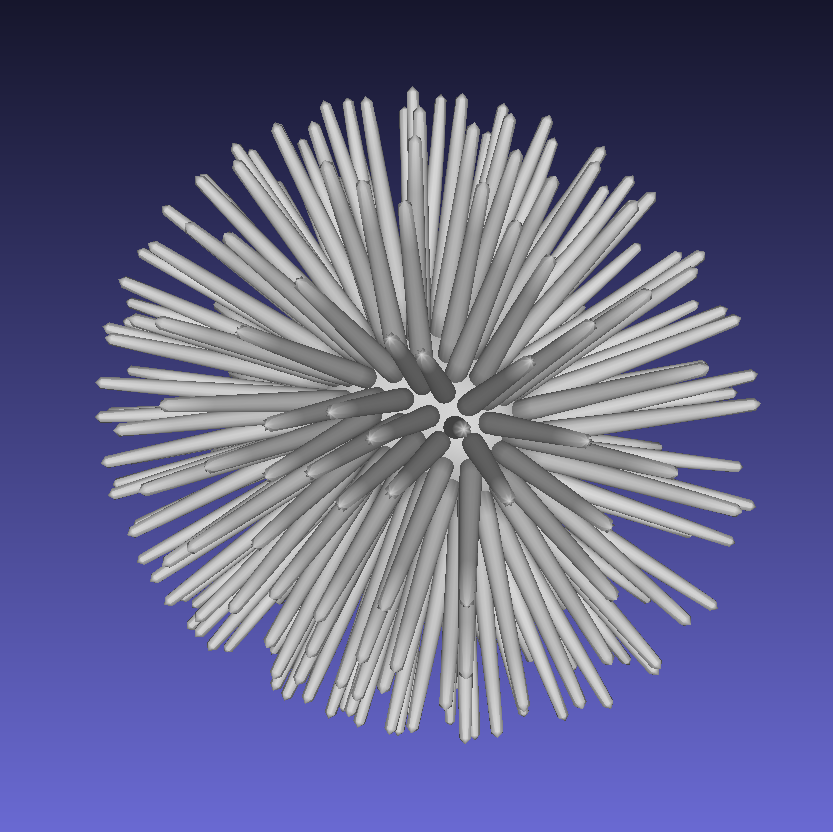}}
     \caption{Polygonal models for testing the quality of the algorithm.}
     \label{fig:polygonal models}
\end{figure}

\begin{figure}[h]
   \centering
   \subbottom[noisy point cloud from the Bunny model]{%
     \includegraphics[width=8cm]{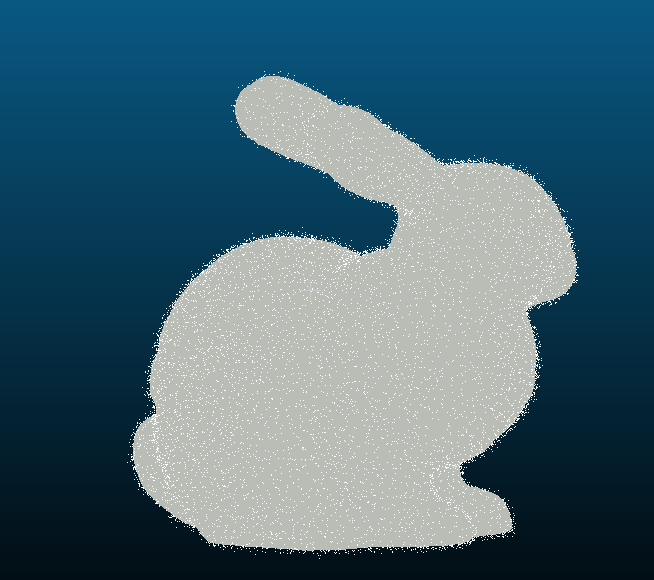}}
   \subbottom[Bunny after MLS]{%
     \includegraphics[width=8cm]{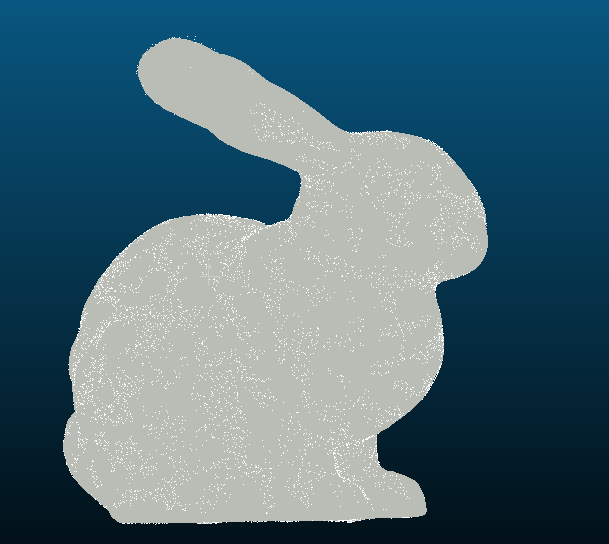}}
   \caption{On the left is a noisy point cloud with additive Gaussian noise with $\boldsymbol{\sigma = 0.01}$ on the reference surface. On the right, a reconstructed MLS point cloud on the same reference surface.}
\end{figure}

\begin{figure}[h]
   \centering
   \subbottom[noisy point cloud from the Bunny model]{%
     \includegraphics[width=0.49\textwidth, height=8cm]{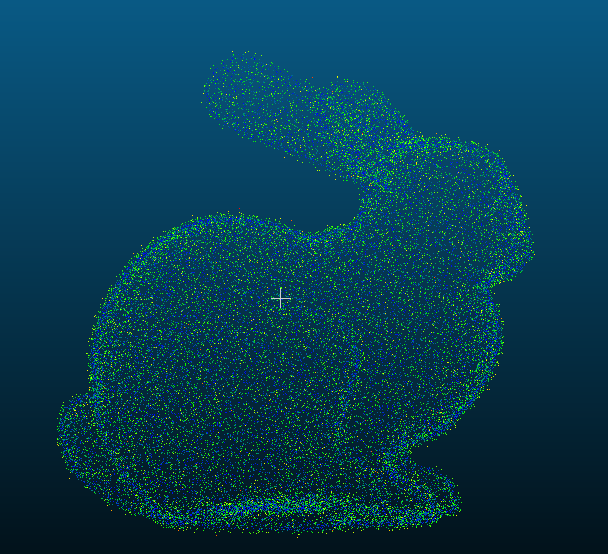}}
   \subbottom[Bunny after MLS]{%
     \includegraphics[width=0.49\textwidth, height=8cm]{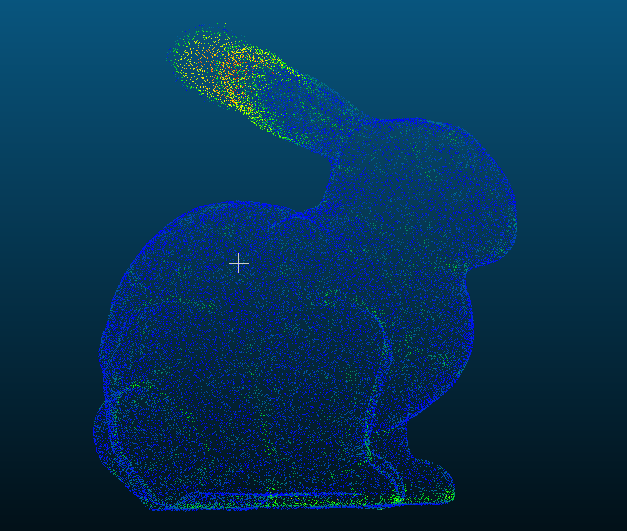}}
   \caption{On the left is a noisy point cloud with additive Gaussian noise with $\boldsymbol{\sigma = 0.01}$. On the right, the reconstructed point cloud using the MLS method. Color intensity indicates the amount of deviation from the reference surface.}
\end{figure}

The tables \ref{table:1}, \ref{table:2}, \ref{table:3}, \ref{table:4}, \ref{table:5}, \ref{table:6} present measurements for various models. The purpose of the measurements was to establish the positive effect of the moving least squares algorithm and to select the optimal free parameter (Neighborhood radius). For all models with the exception of Sea Urchin (Fig. \ref{fig:polygonal models} (f)), it can be seen that the radius value must be selected slightly larger than the noise standard deviation value. This is a good starting estimate given the wide variety of methods for estimating noise distribution. A more accurate parameter value can be selected based on a visual assessment using, for example, the slider button. It is also worth noting for most models the presence of a local minimum up to the standard deviation of the noise. This is due to the fact that the MLS algorithm removes from the result those points for which less than 3 neighbors were found, considering such points to be outliers. When selecting a sufficiently small radius, this leads to significant data loss.
\begin{table}[h]
\centering
\begin{tabular}{| c | c | c| c | c | c | c |}
     \hline
     applied MLS & $\sigma$ & $\bold{R}$ & avg. geom. off & cf. sq. off &min&max\\
     \hline\hline
     $\times$ & 0.005 & -- & 0.00484 & 0.00269 & 3.8e-05 & 0.0245\\
     \checkmark & 0.005 & 0.005 & 0.00372 & 0.00181 & 5.5e-05 & 0.01418\\
     \checkmark & 0.005 & 0.01 & 0.00434 & 0.00245 & 6.7e-05 & 0.01715\\
     \checkmark & 0.005 & 0.03 & \textbf{0.00248} & \textbf{0.00117} & 3.5e-05 & 0.01676\\
     \checkmark & 0.005 & 0.05 & 0.00277 & 0.00185 & 1.3e-05 & 0.02362\\
     \hline
     $\times$ & 0.01 & -- & 0.00854 & 0.00563 & 0.0001 & 0.04387\\
     \checkmark & 0.01 & 0.005 & 0.00578 & 0.00339 & 6.2e-05 & 0.02481\\
     \checkmark & 0.01 & 0.01 & 0.00741 & 0.0045 & 8.7e-05 & 0.03055\\
     \checkmark & 0.01 & 0.03 & \textbf{0.00334} & \textbf{0.002} & 9e-06 & 0.04387\\
     \checkmark & 0.01 & 0.05 & 0.00344 & 0.00226 & 6.1e-05 & 0.03942\\
     \hline
     $\times$ & 0.03 & -- & 0.02333 & 0.01722 & 0.000167 & 0.13393\\
     \checkmark & 0.03 & 0.005 & 0.01424 & 0.01001 & 0.000167 & 0.06133\\
     \checkmark & 0.03 & 0.01 & 0.01672 & 0.01168 & 0.000158 & 0.07857\\
     \checkmark & 0.03 & 0.03 & 0.02065 & 0.01572 & 0.00012 & 0.10811\\
     \checkmark & 0.03 & 0.05 & \textbf{0.01284} & \textbf{0.01116} & 8.3e-05 & 0.11659\\
     \hline
\end{tabular}

\caption{The result of the MLS algorithm on the Bunny model. The input of the algorithm is a cloud of points from the model, noisy with additive Gaussian noise. The reconstructed MLS surface with different parameters R is compared with the reference surface in terms of mean distance, standard deviation, minimum (min) and maximum (max) deviation. $\bold{R}$ is a parameter of the MLS algorithm, $\sigma$ is the standard deviation of additive Gaussian noise.}
\label{table:1}
\end{table}

\begin{table}[h]
\centering
\begin{tabular}{| c | c | c| c | c | c | c |}
     \hline
     applied MLS & $\sigma$ & $\bold{R}$ & avg. geom. off & cf. sq. off &min&max\\
     \hline\hline
     $\times$ & 0.005 & -- & 0.00528 & 0.00266 & 9e-05 & 0.0231\\
     \checkmark & 0.005 & 0.005 & 0.00394 & 0.00176 & 0.000173 & 0.01341\\
     \checkmark & 0.005 & 0.01 & 0.00481 & 0.00229 & 6.1e-05 & 0.01872\\
     \checkmark & 0.005 & 0.03 & \textbf{0.00312} & \textbf{0.00143} & 5.6e-05 & 0.01069\\
     \checkmark & 0.005 & 0.05 & 0.00323 & 0.00149 & 2.1e-05 & 0.01071\\
     \hline
     $\times$ & 0.01 & -- & 0.00896 & 0.00553 & 6.3e-05 & 0.04574\\
     \checkmark & 0.01 & 0.005 & 0.00608 & 0.00321 & 0.000239 & 0.02309\\
     \checkmark & 0.01 & 0.01 & 0.00722 & 0.00395 & 7e-05 & 0.03011\\
     \checkmark & 0.01 & 0.03 & 0.00423 & 0.00242 & 7.1e-05 & 0.04506\\
     \checkmark & 0.01 & 0.05 & \textbf{0.00363} & \textbf{0.00172} & 3e-05 & 0.01496\\
     \hline
     $\times$ & 0.03 & -- & 0.02413 & 0.0175 & 5.5e-05 & 0.13091\\
     \checkmark & 0.03 & 0.005 & 0.01506 & 0.01012 & 0.000453 & 0.06424\\
     \checkmark & 0.03 & 0.01 & 0.01603 & 0.01086 & 0.00022 & 0.08114\\
     \checkmark & 0.03 & 0.03 & 0.02172 & 0.01601 & 7.2e-05 & 0.10536\\
     \checkmark & 0.03 & 0.05 & \textbf{0.0139} & \textbf{0.0122} & 0.000107 & 0.1205\\
     \hline
\end{tabular}

\caption{The result of the MLS algorithm on the Cube model. The input of the algorithm is a cloud of points from the model, noisy with additive Gaussian noise. The reconstructed MLS surface with different parameters R is compared with the reference surface in terms of mean distance, standard deviation, minimum (min) and maximum (max) deviation. $\bold{R}$ is a parameter of the MLS algorithm, $\sigma$ is the standard deviation of additive Gaussian noise.}
\label{table:2}
\end{table}

\begin{table}[h]
\centering
\begin{tabular}{| c | c | c| c | c | c | c |}
     \hline
     applied MLS & $\sigma$ & $\bold{R}$ & avg. geom. off & cf. sq. off &min&max\\
     \hline\hline
     $\times$ & 0.005 & -- & 0.00477 & 0.00268 & 6.1e-05 & 0.02512\\
     \checkmark & 0.005 & 0.005 & 0.00368 & 0.00181 & 6.1e-05 & 0.01509\\
     \checkmark & 0.005 & 0.01 & 0.00424 & 0.00244 & 5.8e-05 & 0.0187\\
     \checkmark & 0.005 & 0.03 & \textbf{0.00265} & \textbf{0.00151} & 4.6e-05 & 0.01855\\
     \checkmark & 0.005 & 0.05 & 0.00301 & 0.00236 & 2.2e-05 & 0.02479\\
     \hline
     $\times$ & 0.01 & -- & 0.00836 & 0.00559 & 0.000109 & 0.04483\\
     \checkmark & 0.01 & 0.005 & 0.00574 & 0.00343 & 0.00012 & 0.02845\\
     \checkmark & 0.01 & 0.01 & 0.00731 & 0.0045 & 5.5e-05 & 0.03267\\
     \checkmark & 0.01 & 0.03 & \textbf{0.00347} & \textbf{0.00222} & 3.8e-05 & 0.04305\\
     \checkmark & 0.01 & 0.05 & 0.00372 & 0.00268 & 6e-05 & 0.03384\\
     \hline
     $\times$ & 0.03 & -- & 0.02264 & 0.01693 & 9.1e-05 & 0.1461\\
     \checkmark & 0.03 & 0.005 & 0.01384 & 0.01005 & 0.000269 & 0.07674\\
     \checkmark & 0.03 & 0.01 & 0.01659 & 0.01172 & 0.000124 & 0.08642\\
     \checkmark & 0.03 & 0.03 & 0.02005 & 0.01549 & 8.6e-05 & 0.10592\\
     \checkmark & 0.03 & 0.05 & \textbf{0.01301} & \textbf{0.01142} & 4.4e-05 & 0.12447\\
     \hline
\end{tabular}

\caption{The result of the MLS algorithm on the Elephant model. The input of the algorithm is a cloud of points from the model, noisy with additive Gaussian noise. The reconstructed MLS surface with different parameters R is compared with the reference surface in terms of mean distance, standard deviation, minimum (min) and maximum (max) deviation. $\bold{R}$ is a parameter of the MLS algorithm, $\sigma$ is the standard deviation of additive Gaussian noise.}
\label{table:3}
\end{table}

\begin{table}[h]
\centering
\begin{tabular}{| c | c | c| c | c | c | c |}
     \hline
     applied MLS & $\sigma$ & $\bold{R}$ & avg. geom. off & cf. sq. off &min&max\\
     \hline\hline
     $\times$ & 0.005 & -- & 0.00477 & 0.00271 & 7.4e-05 & 0.02114\\
     \checkmark & 0.005 & 0.005 & 0.00366 & 0.00181 & 5.3e-05 & 0.01381\\
     \checkmark & 0.005 & 0.01 & 0.00422 & 0.00247 & 5e-05 & 0.01848\\
     \checkmark & 0.005 & 0.03 & \textbf{0.00234} & \textbf{0.00112} & 4e-05 & 0.01742\\
     \checkmark & 0.005 & 0.05 & 0.00243 & 0.00146 & 3.6e-05 & 0.01932\\
     \hline
     $\times$ & 0.01 & -- & 0.00846 & 0.00564 & 0.000118 & 0.04645\\
     \checkmark & 0.01 & 0.005 & 0.00579 & 0.00346 & 0.000106 & 0.02456\\
     \checkmark & 0.01 & 0.01 & 0.00742 & 0.00458 & 0.000103 & 0.03196\\
     \checkmark & 0.01 & 0.03 & \textbf{0.00287} & \textbf{0.00177} & 5e-05 & 0.0449\\
     \checkmark & 0.01 & 0.05 & 0.00299 & 0.00182 & 3.9e-05 & 0.02826\\
     \hline
     $\times$ & 0.03 & -- & 0.02364 & 0.01744 & 0.000117 & 0.1768\\
     \checkmark & 0.03 & 0.005 & 0.01462 & 0.01046 & 0.000183 & 0.07097\\
     \checkmark & 0.03 & 0.01 & 0.01719 & 0.01197 & 8.8e-05 & 0.0872\\
     \checkmark & 0.03 & 0.03 & 0.0208 & 0.01593 & 3.7e-05 & 0.1069\\
     \checkmark & 0.03 & 0.05 & \textbf{0.01255} & \textbf{0.01104} & 0.0001 & 0.12441\\
     \hline
\end{tabular}

\caption{The result of the MLS algorithm on the Hippo model. The input of the algorithm is a cloud of points from the model, noisy with additive Gaussian noise. The reconstructed MLS surface with different parameters R is compared with the reference surface in terms of mean distance, standard deviation, minimum (min) and maximum (max) deviation. $\bold{R}$ is a parameter of the MLS algorithm, $\sigma$ is the standard deviation of additive Gaussian noise.}
\label{table:4}
\end{table}

\begin{table}[h]
\centering
\begin{tabular}{| c | c | c| c | c | c | c |}
     \hline
     applied MLS & $\sigma$ & $\bold{R}$ & avg. geom. off & cf. sq. off &min&max\\
     \hline\hline
     $\times$ & 0.005 & -- & 0.00624 & 0.00275 & 7.3e-05 & 0.02333\\
     \checkmark & 0.005 & 0.005 & \textbf{0.00445} & \textbf{0.00185} & 9.3e-05 & 0.0123\\
     \checkmark & 0.005 & 0.01 & 0.00526 & 0.0022 & 6.9e-05 & 0.01588\\
     \checkmark & 0.005 & 0.03 & 0.00562 & 0.00254 & 9e-05 & 0.02307\\
     \checkmark & 0.005 & 0.05 & 0.00894 & 0.00436 & 0.00017 & 0.02779\\
     \hline
     $\times$ & 0.01 & -- & 0.00956 & 0.00486 & 0.000123 & 0.04262\\
     \checkmark & 0.01 & 0.005 & \textbf{0.00683} & \textbf{0.00305} & 0.000456 & 0.01841\\
     \checkmark & 0.01 & 0.01 & 0.00752 & 0.00341 & 0.000226 & 0.02633\\
     \checkmark & 0.01 & 0.03 & 0.00863 & 0.0045 & 5.9e-05 & 0.03821\\
     \checkmark & 0.01 & 0.05 & 0.01034 & 0.00485 & 7e-05 & 0.04272\\
     \hline
     $\times$ & 0.03 & -- & 0.01812 & 0.01358 & 0.000175 & 0.1198\\
     \checkmark & 0.03 & 0.005 & \textbf{0.01033} & \textbf{0.00584} & 0.000688 & 0.05006\\
     \checkmark & 0.03 & 0.01 & 0.01103 & 0.00631 & 0.000276 & 0.05743\\
     \checkmark & 0.03 & 0.03 & 0.01652 & 0.01145 & 0.000153 & 0.09185\\
     \checkmark & 0.03 & 0.05 & 0.01624 & 0.01225 & 0.000219 & 0.09817\\
     \hline
\end{tabular}

\caption{The result of the MLS algorithm on the Sea Urchin model. The input of the algorithm is a cloud of points from the model, noisy with additive Gaussian noise. The reconstructed MLS surface with different parameters R is compared with the reference surface in terms of mean distance, standard deviation, minimum (min) and maximum (max) deviation. $\bold{R}$ is a parameter of the MLS algorithm, $\sigma$ is the standard deviation of additive Gaussian noise.}
\label{table:5}
\end{table}

\begin{table}[h]
\centering
\begin{tabular}{| c | c | c| c | c | c | c |}
     \hline
     applied MLS & $\sigma$ & $\bold{R}$ & avg. geom. off & cf. sq. off &min&max\\
     \hline\hline
     $\times$ & 0.005 & -- & 0.00486 & 0.00269 & 6.7e-05 & 0.0226\\
     \checkmark & 0.005 & 0.005 & 0.00371 & 0.00179 & 7.6e-05 & 0.01491\\
     \checkmark & 0.005 & 0.01 & 0.00437 & 0.00245 & 8.7e-05 & 0.01768\\
     \checkmark & 0.005 & 0.03 & \textbf{0.00248} & \textbf{0.00113} & 3.8e-05 & 0.01193\\
     \checkmark & 0.005 & 0.05 & 0.00256 & 0.00125 & 3.1e-05 & 0.01736\\
     \hline
     $\times$ & 0.01 & -- & 0.00856 & 0.00564 & 8.3e-05 & 0.0453\\
     \checkmark & 0.01 & 0.005 & 0.0058 & 0.00336 & 0.000122 & 0.0258\\
     \checkmark & 0.01 & 0.01 & 0.00741 & 0.00447 & 3.8e-05 & 0.03209\\
     \checkmark & 0.01 & 0.03 & 0.00329 & 0.00186 & 3e-05 & 0.04396\\
     \checkmark & 0.01 & 0.05 & \textbf{0.00313} & \textbf{0.0016} & 3.4e-05 & 0.02138\\
     \hline
     $\times$ & 0.03 & -- & 0.02359 & 0.01737 & 7.6e-05 & 0.14377\\
     \checkmark & 0.03 & 0.005 & 0.01461 & 0.01033 & 0.000214 & 0.06379\\
     \checkmark & 0.03 & 0.01 & 0.01675 & 0.01164 & 6.5e-05 & 0.09195\\
     \checkmark & 0.03 & 0.03 & 0.0208 & 0.01587 & 0.000106 & 0.10975\\
     \checkmark & 0.03 & 0.05 & \textbf{0.0124} & \textbf{0.01011} & 3.2e-05 & 0.12571\\
     \hline
\end{tabular}

\caption{The result of the MLS algorithm on the Woman model. The input of the algorithm is a cloud of points from the model, noisy with additive Gaussian noise. The reconstructed MLS surface with different parameters R is compared with the reference surface in terms of mean distance, standard deviation, minimum (min) and maximum (max) deviation. $\bold{R}$ is a parameter of the MLS algorithm, $\sigma$ is the standard deviation of additive Gaussian noise.}
\label{table:6}
\end{table}
\section*{Approbation}
\addcontentsline{toc}{section}{Approbation}
M.I. Khabibulin. Study of the effectiveness of the moving least squares method in reconstructing a three-dimensional surface on a supercomputer. Conference "Lomonosov": proceedings of the international scientific conference "Lomonosov". April 10-21, 2023, pp. 40-43, Moscow.

\section*{Main results}
\addcontentsline{toc}{section}{Main results}
\begin{itemize}
     \item The most popular methods of surface representation and reconstruction are analyzed.
     \item The following variants of the parallel surface reconstruction algorithm have been implemented:
for systems with shared memory (using OpenMP), with distributed memory (using MPI),
as well as a hybrid version (MPI + OpenMP)
     \item The effectiveness of the developed algorithm was studied, and tests of the reconstruction of real surfaces were carried out.
     \item Computational experiments have shown the effectiveness of the developed implementations
\end{itemize}

\section*{Conclusions and conclusion}
\addcontentsline{toc}{section}{Conclusions and conclusion}
The work discusses reconstruction methods that represent the surface in various ways. Delaunay triangulation represents a surface with a polygonal mesh. The radial basis function method represents a surface as a set of implicit functions.
The local optimal projection operator and the moving least squares method are intermediate options and represent a surface as a set of points. Both algorithms can also upscale pixel density to screen resolution and be used for subsequent rendering. Quite often, algorithms that represent a surface as a set of points are used to obtain an intermediate result and then apply algorithms that represent the surface as a polygonal mesh or a set of implicit functions. Today, representing a surface with a polygonal mesh is the most used. Although the authors of ~\cite{CARR} ~\cite{Turk} claim that representing a surface with implicit functions has wide application, today, representing a surface with a polygonal mesh has supplanted this approach. This is evidenced by the lack of ability to render complex surfaces represented by implicit functions on the most popular rendering software (Unity, Blender, etc.).

The work formulates and implements a parallel surface reconstruction algorithm using the Message Passing Interface software interface for message transmission, based on the moving least squares algorithm. The algorithm provides for uniform distribution of point cloud segments among processes with subsequent forwarding of partition boundaries along the ring topology. All subsequent calculations are performed locally. By eliminating subsequent data transfers, it provides maximum speedup.

The algorithm also has a large parallelism resource due to the nonlinear decrease in the number of operations at some stages of the algorithm with a linear increase in the number of processes (see table ~\ref{table:complexity}). This applies to the stages of constructing a k-d-tree and finding neighbors for all points of the segment distributed to the process.

According to research results, the MLS algorithm copes well with noisy data, but to achieve optimal results, you need to carefully select the radius parameter of the algorithm. Another advantage of the algorithm is that there is only one free parameter R (the search radius for neighboring points) versus three for the Delaunay triangulation. As a direction for further development of the algorithm, it is interesting to formulate guidelines for choosing the optimal parameter of the algorithm and automate this process.

    \clearpage
    \printbibliography
\end{document}